\documentclass[12pt]{article}

\usepackage{epsfig} 
\usepackage{amsbsy}

\def\beq{\begin{equation}}
\def\eeq{\end{equation}}

\def\beqn{\begin{eqnarray}}
\def\eeqn{\end{eqnarray}}

\def\lq{\left[}
\def\rq{\right]}
\def\rg{\right\}}
\def\lg{\left\{}
\def\({\left(}
\def\){\right)}

\def\Dx{\int {\cal D}x }
\def\I{{\cal I}}
\def\Q{{\cal Q} }
\def\P{{\cal P} }
\def\X{{\cal X} }
\def\Y{{\cal Y} }

\newcommand{\G}[1]{\Gamma\left({#1}\right)}
\newcommand{\n}[1]{\nu_{#1}}
\newcommand{\halfD}{\frac{D}{2}}

\def\k#1{k_{#1}}
\def\K#1{K_{#1}}

\def\p#1{\nu_{#1}{\bf #1^{\boldsymbol{+}}}}
\def\ps#1{{\bf #1^{\boldsymbol{+}}}}
\def\m#1{{\bf #1^{\boldsymbol{-}}}}

\def\Jd{J^D\(\{\nu_i\};\{Q_i^2\} \)}

\def\Pbox{J^D\(\nu_1,\nu_2,\nu_3,\nu_4,\nu_5,\nu_6,\nu_7;s,t\)}
\def\Abox{A^D\(\nu_1,\nu_2,\nu_3,\nu_5,\nu_7;s,t\)}
\def\Bbox{B^D\(\nu_2,\nu_3,\nu_4,\nu_6,\nu_7;s,t\)}
\def\Cbox{C^D\(\nu_2,\nu_3,\nu_5,\nu_6,\nu_7;s,t\)}

\def\BUB{\Pi^D}
\def\pow{(-1)^\halfD~}

\def\measure#1{ \int \frac{d^D \k#1}{i \pi^{D/2}}}
\def\Measure#1{ \int \frac{d^D \K#1}{i \pi^{D/2}}}

\def\dplus{{\bf d^{\boldsymbol{+}}}} 
\def\dminus{{\bf d^{\boldsymbol{-}}}} 

\def\f21{{_2F_1}}
\def\ftt{{_3F_2}}

\def\fracmut{\frac{s+t}{t}}
\def\fracmtu{\frac{t}{s+t}}

\def\fracmus{\frac{s+t}{s}}
\def\fracmsu{\frac{s}{s+t}}

\def\al{\alpha}

\def\li#1{\,{\rm Li}_2\left(#1 \right)}
\def\lit#1{\,{\rm Li}_3\left(#1 \right)}
\def\lif#1{\,{\rm Li}_4\left(#1 \right)}
\def\sot#1{\,{\rm S}_{1,\,2}\left(#1 \right)}
\def\so3#1{\,{\rm S}_{1,\,3}\left(#1 \right)}
\def\st2#1{\,{\rm S}_{2,\,2}\left(#1 \right)}

\def\ss#1#2#3{{\rm S}_{#1,\,#2}\(#3\)}

\newskip\humongous \humongous=0pt plus 1000pt minus 1000pt

\newif\ifdtup

\catcode`\@=11

\@addtoreset{equation}{section}

\def\theequation{\thesection.\arabic{equation}}

\def\@normalsize{\@setsize\normalsize{15pt}\xiipt\@xiipt
\abovedisplayskip 14pt plus3pt minus3pt%
\belowdisplayskip \abovedisplayskip
\abovedisplayshortskip \z@ plus3pt%
\belowdisplayshortskip 7pt plus3.5pt minus0pt}

\def\small{\@setsize\small{13.6pt}\xipt\@xipt
\abovedisplayskip 13pt plus3pt minus3pt%
\belowdisplayskip \abovedisplayskip
\abovedisplayshortskip \z@ plus3pt%
\belowdisplayshortskip 7pt plus3.5pt minus0pt
\def\@listi{\parsep 4.5pt plus 2pt minus 1pt
     \itemsep \parsep
     \topsep 9pt plus 3pt minus 3pt}}

\@twosidetrue

\catcode`@=12

\catcode`\@=11

\def\section{\@startsection{section}{1}{\z@}{3.5ex plus 1ex minus
   .2ex}{2.3ex plus .2ex}{\large\bf}}

\def\thesection{\arabic{section}}
\def\thesubsection{\arabic{section}.\arabic{subsection}}
\def\thesubsubsection{\arabic{section}.\arabic{subsection}.\arabic{subsubsection}}

\def\appendix{\setcounter{section}{0}
 \def\thesection{\Alph{section}}
 \def\theequation{\Alph{section}.\arabic{equation}}
 \def\thesubsection{\Alph{section}.\arabic{subsection}}
\def\thesubsubsection{\Alph{section}.\arabic{subsection}.\arabic{subsubsection}}

 \def\section{\@startsection{section}{1}{\z@}{3.5ex plus 1ex minus
   .2ex}{2.3ex plus .2ex}{\large\bf}}
}

\newcommand{\ccaption}[2]{
  \begin{center}
    \parbox{0.85\textwidth}{
      \caption[#1]{\small\it {#2}}}
  \end{center}    }

\def \ep{\epsilon}
\def \to   {\mbox{$\rightarrow$}}

\newcommand\hepph[1]{{\tt hep-ph/#1}}

\def\ord#1{{\cal O}\(#1\)}

\begin{document}
\begin{titlepage}
\nopagebreak

%

{\flushright{
        \begin{minipage}{4cm}
         DTP/99/106 \\
        {\tt hep-ph/9912251}\hfill \\
        \end{minipage}        }

}
\vfill
\begin{center}
{\LARGE \bf \sc
 \baselineskip 0.9cm
The two-loop scalar and tensor 
Pentabox graph with light-like legs
    
}
\vskip
1.3cm 
{\large  C.~Anastasiou\footnote{e-mail: {\tt Ch.Anastasiou@durham.ac.uk}},
E.~W.~N.~Glover\footnote{e-mail: {\tt E.W.N.Glover@durham.ac.uk}}  and
C.~Oleari\footnote{e-mail: {\tt Carlo.Oleari@durham.ac.uk}}} 
\vskip .2cm 
{\it Department of Physics, 
University of Durham, 
Durham DH1 3LE, 
England } 
\vskip
1.3cm    
\end{center}

\nopagebreak
\begin{abstract}
We study the scalar and tensor integrals associated with the {\bf pentabox}
topology: the class of two-loop box integrals with seven propagators - five
in one loop and three in the other.  We focus on the case where the external
legs are light-like and use integration-by-parts identities to express the
scalar integral in terms of two master-topology integrals and present an
explicit analytic expression for the pentabox scalar integral as a series
expansion in $\ep = (4-D)/2$.  We also give an algorithm based on integration
by parts for relating the generic tensor integrals to the same two master
integrals and provide general formulae describing the master integrals in
arbitrary dimension and with general powers of propagators.
\end{abstract}
\vfill
\vfill
\end{titlepage}
\newpage                                                                     
\section{Introduction}
\label{sec:intro}

Scattering processes are one of the most important sources of information on
short distance physics and have played a vital role in establishing the
properties of the fundamental interactions of nature.  At the one-loop level, box
integrals are a key ingredient in the comparison of perturbative predictions
with experimental data, such as wide-angle jet
production in hadron collisions or Bhabha scattering in electron-positron
annihilation.  Recent improvements of experimental measurements demand even
more precise theoretical predictions and there is significant interest in
determining $2 \,\to\, 2$ scattering rates at the two-loop order.  Achieving
this goal requires the evaluation of  two-loop graphs such as
the planar double-box graph~\cite{Smirnov,Smirnov2}, the non-planar
double-box graph~\cite{Bas} or some one-loop box integrals with bubble
insertions on one of the propagators~\cite{AGO2}. 

\begin{figure}[t]
\begin{center}
\epsfig{file=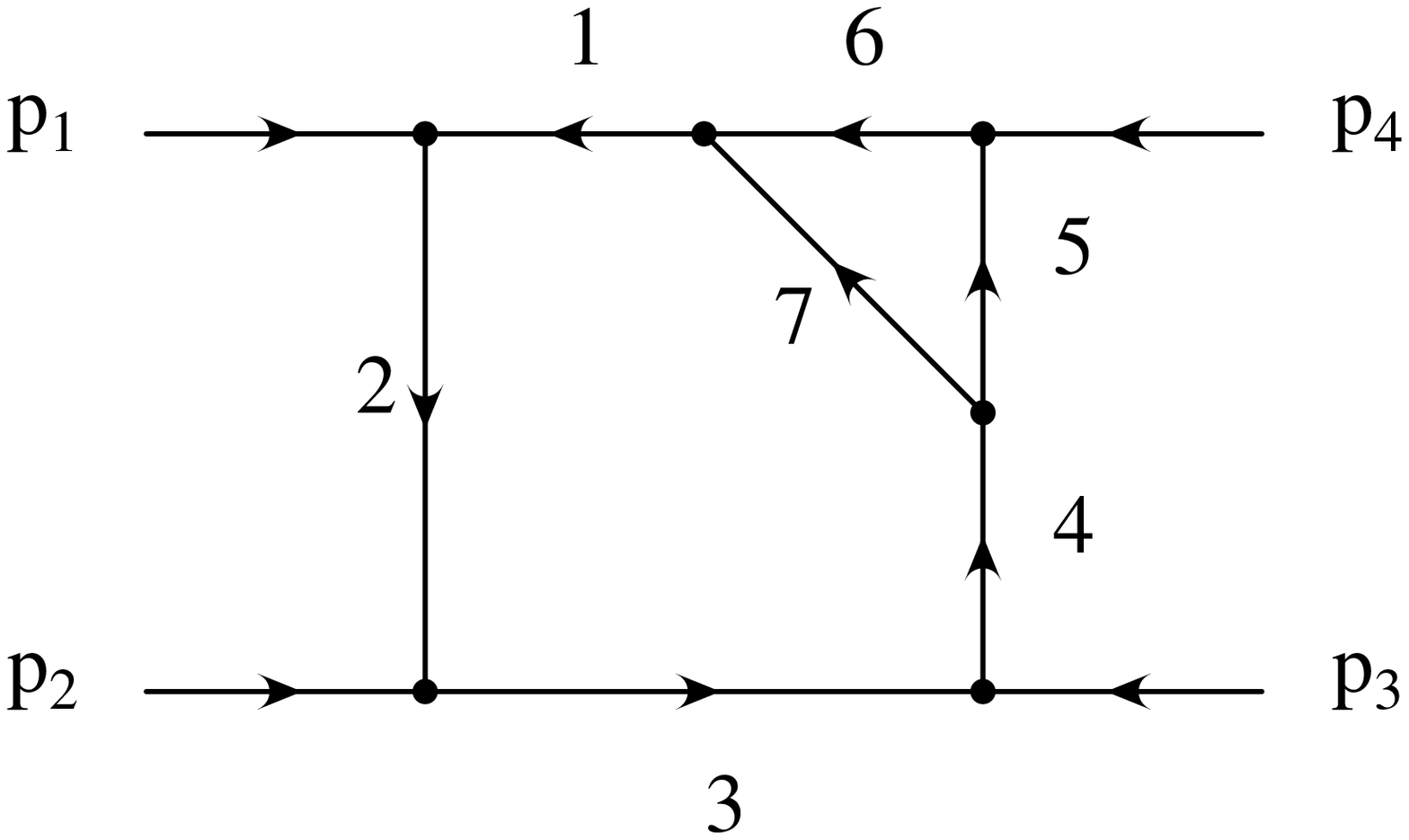,height=5cm}
\end{center}
\label{fig:pbox}
\ccaption{}{The pentabox topology.   The propagators are labelled according to
Eq.~(\ref{eq:props}) and are each raised to the $\nu_i$ power.}
\end{figure}

In addition, we also need to study the particular type of planar box graph
shown in Fig.~\ref{fig:pbox}.  Owing to the particular shape of this diagram,
we name this topology {\bf pentabox}. It is the purpose of this paper to
provide analytic expressions for the scalar pentabox with unit powers of the
propagators using dimensional regularisation for the light-like (or on-shell)
case, $p_1^2=p_2^2=p_3^2=p_4^2=0$, in terms of generalised polylogarithms.
This type of integral is much simpler than the planar
double-box~\cite{Smirnov} and the non-planar double-box~\cite{Bas} because of
the presence of the triangle sub-graph which allows the use of
integration-by-parts identities~\cite{ibyp} to relate the most general scalar
integral to two types of simpler integrals with fewer propagators, which we
denote {\em master topologies}.  We give expressions for these simpler
integrals and use them to derive an analytic expression for the pentabox
graph with unit propagators as a series expansion in $\epsilon = (4-D)/2$.
We also give algorithms for evaluating all the associated tensor integrals.
These are directly related to scalar integrals with higher powers of the
propagators in higher dimension.  As with the scalar integral, the
integration-by-parts identities allow the tensor integrals to be obtained in
terms of the two master-topology integrals.

Our paper is organised as follows.  In Section~\ref{sec:notation} we
establish the basic notation we will use for two-loop integrals.  Analytic
expressions for the scalar pentabox integral, together with algorithms
for reducing the integral to the simpler master topologies, are given in
Section~\ref{sec:scalar}, while the tensor integrals are described in
Section~\ref{sec:tensor}.  Details of the master topologies and algorithms
relating them are given in Section~\ref{sec:master}.  Explicit expansions
in $\epsilon = (4-D)/2$ are computed in terms of generalised polylogarithm
functions.  Useful relations amongst the polylogarithms are collected in
Appendix~\ref{sec:app}.  Finally, our findings are summarized in
Section~\ref{sec:conc}.

\section{Notation}
\label{sec:notation}

We denote the generic two-loop integral in $D$ dimensions with 
$n$ propagators $A_i$ raised to
arbitrary powers as
\begin{equation}
\Jd\lq 1;\,\k1^\mu;\,\k2^\mu;\,\k1^\mu\k1^\nu;\,\k1^\mu\k2^\nu;\,\ldots \rq
=
\measure1
\measure2
\frac{\lq 1;\,\k1^\mu;\,\k2^\mu;\,\k1^\mu\k1^\nu;\,\k1^\mu\k2^\nu;\,\ldots 
\rq}{ A_1^{\nu_1}\cdots A_n^{\nu_n}},
\label{eq:Jdef}
\end{equation}
where the external momentum scales are indicated by $\{Q_i^2\}$.  When the
numerator in Eq.~(\ref{eq:Jdef}) is unity, $\Jd[1] \equiv \Jd$, we have the
scalar integral that forms the basic underlying skeleton of the graph. The
tensor integrals $\Jd[\k1^\mu;\, \ldots]$, which naturally arise in Feynman
diagrams, are usually related to combinations of the scalar integral with
different powers of propagators and/or different values of $D$ \cite{Tar}.
Evaluation of two-loop matrix elements for physical processes requires a
knowledge of both tensor and scalar integrals.

In this paper we focus on the specific two-loop box graph shown in
Fig.~\ref{fig:pbox}, whose analytic expression is
\begin{equation}
\Pbox
=
\measure1
\measure2
~
\frac{1}{
A_1^{\nu_1}
A_2^{\nu_2}
A_3^{\nu_3}
A_4^{\nu_4}
A_5^{\nu_5}
A_6^{\nu_6}
A_7^{\nu_7}},
\end{equation}
together with the associated tensor integrals, where the propagators are
given by
\begin{eqnarray}
\label{eq:props}
A_1 &=& \k1^2 + i0, \nonumber \\
A_2 &=& (\k1+p_1)^2 + i0, \nonumber \\
A_3 &=& (\k1+p_1+p_2)^2 + i0, \nonumber \\
A_4 &=& (\k1+p_1+p_2+p_3)^2 + i0, \\
A_5 &=& (\k2+p_1+p_2+p_3)^2 + i0, \nonumber \\
A_6 &=& \k2^2 + i0, \nonumber \\
A_7 &=& (\k1-\k2)^2 + i0. \nonumber
\end{eqnarray}
With this momentum assignment we see that the loop momentum $\k1$ circulates
in five of the propagators while $\k2$ circulates in only three.  All of the
external momenta are in-going and are taken to be light-like:
$p_1^2=p_2^2=p_3^2=p_4^2=0$.  The only momentum scales present in the problem
are therefore the usual Mandelstam variables $s = (p_1+p_2)^2$ and
$t=(p_2+p_3)^2$, together with $u=-s-t$.

\section{The scalar pentabox integral}
\label{sec:scalar}

The generic scalar pentabox integral, $\Pbox$, with arbitrary powers of
propagators in $D$ dimension is always reducible to a finite set of simpler
master-topology
integrals. In fact, applying the integration-by-parts method \cite{ibyp} to
the following two integrals
\beq
\measure1 \measure2 \;\frac{\partial}{\partial k_2^\mu} \;
\frac{\lq \(k_2-p_4\)^\mu; \; k_2^\mu \rq}{A_1^{\nu_1} A_2^{\nu_2}
A_3^{\nu_3} A_4^{\nu_4} A_5^{\nu_5} A_6^{\nu_6} A_7^{\nu_7}} = 0,
\eeq
we find that
\begin{eqnarray}
\label{eq:IbyPa}
\(D-2\nu_5-\nu_6-\nu_7\) J^D &=&
\left(\p6 \m5+ \p7\m5-\p7\m4\right) J^D, \\  
\label{eq:IbyPb}
\(D-\nu_5-2\nu_6-\nu_7\) J^D &=&
\left(\p5\m6+ \p7\m6-\p7\m1\right) J^D,
\end{eqnarray}
where $J^D=\Pbox$ and the symbols ${\bf i^{\boldsymbol{+}}}$ and ${\bf i^{\boldsymbol{-}}}$ are shorthand
notation for raising and lowering of powers of propagator $i$
\beq
{\bf i^{\boldsymbol{\pm}}} J^D(\ldots, \nu_i, \ldots) = J^D(\ldots,
\nu_i\pm1, \ldots). 
\eeq
As usual, each raising operator is always accompanied by a factor of $\nu_i$
so that it is impossible to raise the power of the propagator if it is not
already present, i.e. $\nu_i \neq 0$.

By repeated application of Eq.~(\ref{eq:IbyPa}), we can reduce either of
$\nu_4$ or $\nu_5$ to zero.  Similarly, by applying Eq.~(\ref{eq:IbyPb})
we can lower (and eventually eliminate) the power of either $\nu_1$ or
$\nu_6$.  This produces a pinched graph of the form
\begin{eqnarray}
\label{eq:master}
J^D\(\nu_1,\nu_2,\nu_3,\nu_4,0,0,\nu_7;s,t\) &=& 0, \\
\label{eq:masterA}
J^D\(\nu_1,\nu_2,\nu_3,0,\nu_5,0,\nu_7;s,t\) &=&
\Abox,\\
J^D\(0,\nu_2,\nu_3,\nu_4,0,\nu_6,\nu_7;s,t\)&=& 
\label{eq:masterB}
\Bbox,\\
\label{eq:masterC}
J^D\(0,\nu_2,\nu_3,0,\nu_5,\nu_6,\nu_7;s,t\)&=&
\Cbox,
\end{eqnarray}
where, diagrammatically,
\begin{eqnarray*}
\hspace{-1cm}
\Abox  &\equiv& \begin{minipage}{7cm}
 \epsfig{file=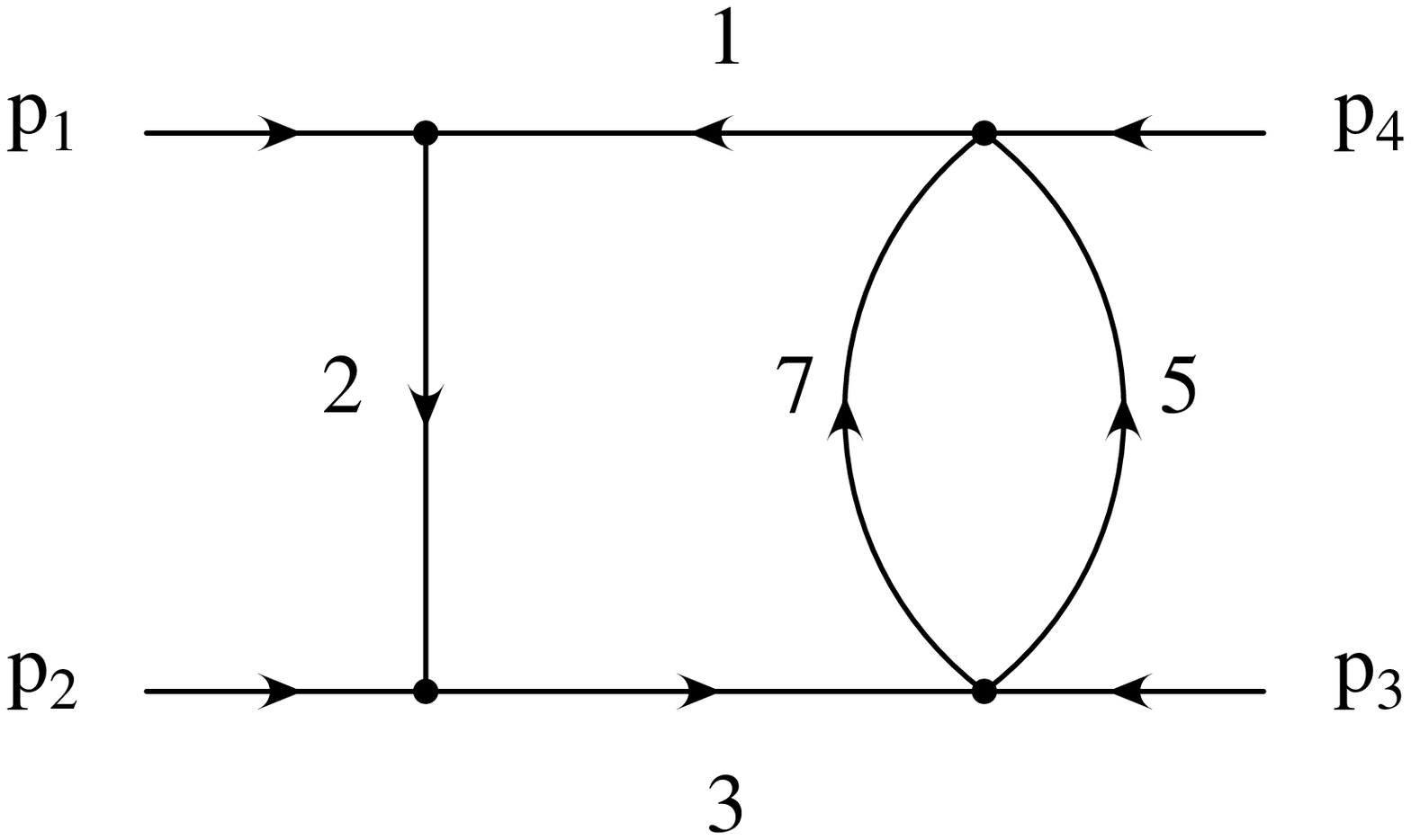,height=4cm}
\end{minipage} 
\\
\Bbox &\equiv& \begin{minipage}{7cm}
 \epsfig{file=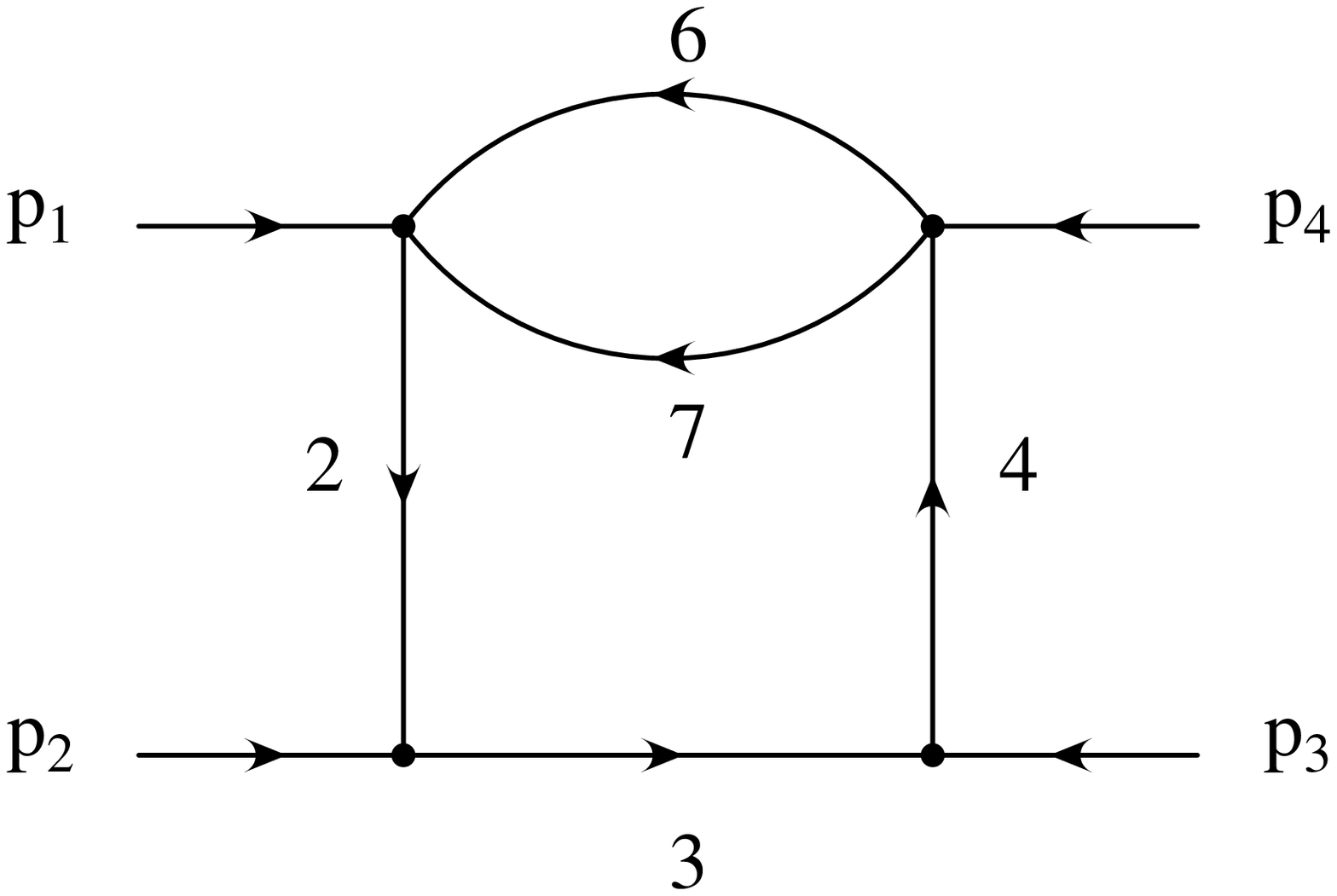,height=4.5cm}
\end{minipage} 
\\
\Cbox &\equiv& \begin{minipage}{7cm}
 \epsfig{file=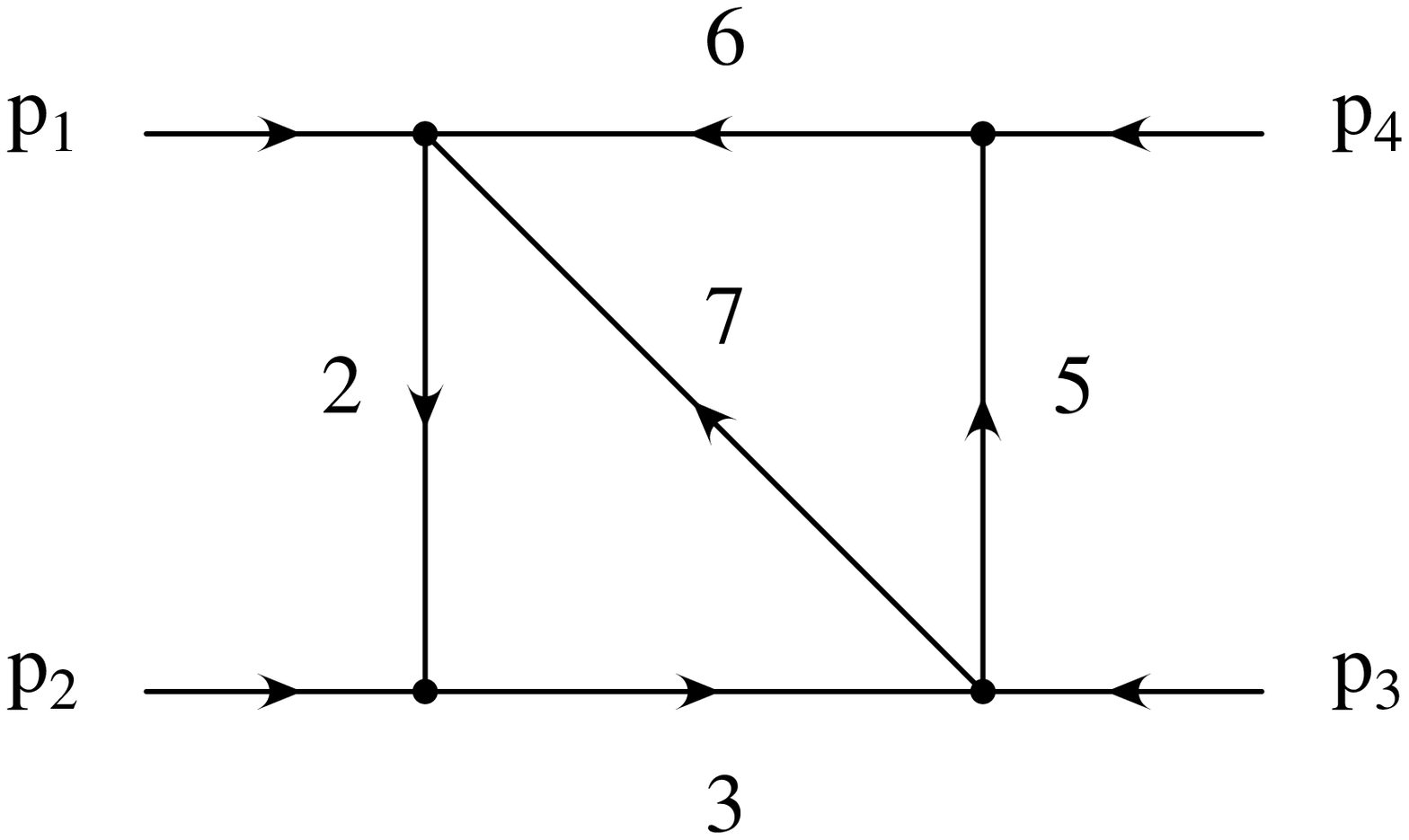,height=4cm}
\end{minipage} 
\end{eqnarray*}
$A^D$, $B^D$ and $C^D$ constitute our set of master-topology integrals.
Due to the particular symmetry of these diagrams, they satisfy some
additional identities:
\begin{itemize}
\item[-] The basic $A^D$ and $B^D$ integrals are related by exchanging $s$
and $t$ according to
\begin{equation}
\label{eq:ABrel}
B^D(\nu_2,\nu_3,\nu_4,\nu_6,\nu_7;s,t)
=
A^D(\nu_2,\nu_3,\nu_4,\nu_6,\nu_7;t,s),
\end{equation}
and $A^D$ has the additional symmetries
\begin{eqnarray}
\label{eq:Asym}
A^D(\nu_3,\nu_2,\nu_1,\nu_5,\nu_7;s,t)&=&
A^D(\nu_1,\nu_2,\nu_3,\nu_5,\nu_7;s,t),\nonumber \\
A^D(\nu_1,\nu_2,\nu_3,\nu_7,\nu_5;s,t)&=&
A^D(\nu_1,\nu_2,\nu_3,\nu_5,\nu_7;s,t).
\end{eqnarray}

\item[-] Similarly, the basic $C^D$ integral has the properties
\begin{eqnarray}
\label{eq:Csym}
C^D(\nu_6,\nu_5,\nu_3,\nu_2,\nu_7;t,s) &=&
C^D(\nu_2,\nu_3,\nu_5,\nu_6,\nu_7;s,t),\nonumber \\
C^D(\nu_3,\nu_2,\nu_6,\nu_5,\nu_7;t,s) &=&
C^D(\nu_2,\nu_3,\nu_5,\nu_6,\nu_7;s,t).
\end{eqnarray}
\end{itemize}

\paragraph{Application:}

We consider here the specific case of the scalar integral with all the
propagators equal to one. Applying Eqs.~(\ref{eq:IbyPa}) and~(\ref{eq:IbyPb})
we obtain
\begin{eqnarray}
\lefteqn{\hspace{-1cm}
J^D\(1,1,1,1,1,1,1;s,t\) =\frac{1}{(D-5)(D-4)}}\nonumber \\
&\times & \lq 2 C^D(1,1,1,1,3;s,t) -2 A^D(1,1,1,1,3;s,t) -
A^D(1,1,1,2,2;s,t)\right. \nonumber \\
&&\left. -2 B^D(1,1,1,1,3;s,t) -
B^D(1,1,1,2,2;s,t) \rq.
\end{eqnarray}
The explicit form for the master-topology integrals with arbitrary powers of
propagators and arbitrary dimension are collected in Section~\ref{sec:master}.
Making a series expansion in $\epsilon = (4-D)/2$, we find that the general
structure of the scalar integral can be written
\beqn
\label{eq:pentaboxts}
J^D\(1,1,1,1,1,1,1;s,t\) &= &
\frac{3}{4\, \ep^4} \frac{(1+3\ep)}{(1+\ep)(1+2\ep)}
\frac{\G{1-\ep}^3 \G{1+2\ep}}{\G{1-3\ep}} 
\nonumber\\
&& {} \times\lq \frac{(-t)^{-2\ep}}{st^2} R(s,t) + \frac{(-s)^{-2\ep}}{ts^2}
R(t,s)  \rq.
\eeqn

\begin{enumerate}
\item[1)]
In the physical region $\boldsymbol{s>0}$, $\boldsymbol{t<0}$ we have
\begin{eqnarray}
\label{eq:pentabox_expr}
R(s,t)&=& 1+\epsilon +  2\zeta_2\,\epsilon^2 + \lq 4 \sot{\frac{s+t}{s}} 
+ 2 \zeta_3 + 6 \zeta_2 \rq \epsilon^3 \nonumber \\
&& \hspace{-3mm} {} +\bigg\{
12 \st2{\frac{s+t}{s}}-8 \so3{\frac{s+t}{s}}+8 \sot{\frac{s+t}{s}}
+28 \zeta_2 \li{\frac{s+t}{s}} \nonumber \\
&& \hspace{-3mm} {} +6 \zeta_3 -51 \zeta_4 \bigg\} \,\epsilon^4 +4 \,i\, \pi
\,\Theta(s, t)  \lg \lq \li{ \frac{s+t}{s} } - \zeta_2\rq 
\ep^3 \right.\nonumber \\ 
&&  \hspace{-3mm} {}+\left. \lq 3 \lit{\frac{s+t}{s}} + 2 
\li{\frac{s+t}{s}}  - 2  \sot{\frac{s+t}{s}}  -\zeta_3- 2 \zeta_2 \rq 
\ep^4 \rg + \ord{\ep^5}.\nonumber \\ 
\end{eqnarray}
The generalised polylogarithms (defined in Section~\ref{subsec:defs}) that
appear in Eq.~(\ref{eq:pentabox_expr}), and then in
Eq.~(\ref{eq:pentaboxts}), have  arguments $(s+t)/t$ and
$(s+t)/s$, which are always less than unity in this region.  The function
$\Theta(t, s)$ assumes values
\beq 
\Theta(s, t) =  - \Theta(t,s) = -1.
\eeq
The imaginary parts associated with the prefactor can be simply obtained by
recalling the $+i0$ prescription associated with the external kinematic
scales
\beqn
(-s)^{-2\ep} &\to& |s|^{-2\ep} \, \exp \(2\pi i \ep\),\\
(-t)^{-2\ep} &\to& |t|^{-2\ep}.
\eeqn 

\item[2)]
In the other physical region $\boldsymbol{s<0}$, $\boldsymbol{t<0}$, we use
the analytic continuation formulae of Section~\ref{subsec:anal}, and we find
\def\sotps{\frac{s}{s+t}}
\def\tpsos{\frac{s+t}{s}}
\beqn
R(s,t)&=& 1+\epsilon +  2\zeta_2\,\epsilon^2 + 
\lq  4\lit{\sotps} +4 \log\(\tpsos\) \li{\sotps}\right. \nonumber\\
&&  \hspace{-3mm} {} 
\left. + \frac{2}{3}\log^3\(\tpsos\)
+2\pi^2\log\(\tpsos\)-4\sot{\sotps} +6\zeta_3+\pi^2  \rq \ep^3 \nonumber\\
&&  \hspace{-3mm} {}
+ \lg 4 \st2{\sotps} + 8 \so3{\sotps} -8 \lq \log\(\tpsos\) +1 \rq
\sot{\sotps} \right. \nonumber\\
&& \hspace{-3mm} {}
 - 16 \lif{\sotps} + 4\lq 2  -   \log\(\tpsos\) \rq \lit{\sotps}
+ \frac{5}{6}\log^4 \(\tpsos\)\nonumber\\
&& \hspace{-3mm} {}
+ 4 \lq \log^2\(\tpsos\) + 2 \log\(\tpsos\) - \zeta_2 \rq \li{\sotps} 
+ \frac{4}{3} \log^3\(\tpsos\) \nonumber\\
&& \hspace{-3mm} {}
+ \left.\frac{8}{3} \pi^2\log^2\(\tpsos\) + 12 \lq \zeta_3 +2\zeta_2 \rq
\log\(\tpsos\) + 14 \zeta_3 + 31 \zeta_4\rg \ep^4 + \ord{\ep^5}.\nonumber\\
\eeqn
Notice that, in this region, the scalar pentabox integral of
Eq.~(\ref{eq:pentaboxts}) has no imaginary part, as expected.
\end{enumerate}
\section{The generic tensor integrals}
\label{sec:tensor}

In this section, we want to deal with the  generic tensor integral,
$\Jd[\k1^\mu;\,\ldots]$, using the pentabox just as an example of how to
reduce the tensor integral to a set of simpler scalar integrals.

A method to reduce tensor integrals constructing differential operators that
change the powers of the propagators as well as the dimension of the integral
was presented in Ref.~\cite{Tar}.  However, it is in our view simpler to
obtain the tensor integrals directly from the Schwinger parameterised form of
the integral.  In fact, in its most general form, the integrand can be
rewritten as
\begin{equation}
\frac{1}{A_1^{\nu_1}\ldots A_n^{\nu_n}}
=  \Dx~\exp\left ( \sum_{i=1}^n x_iA_i \right),
\end{equation}
where
\begin{equation}
\label{eq:Dx}
\Dx =  \prod_{i=1}^n \frac{(-1)^{\nu_i}}{\Gamma(\nu_i)}
\int^{\infty}_0 dx_i\, x_i^{\nu_i-1} 
\end{equation}
and
\begin{equation}
\sum_{i=1}^n x_iA_i = a\, \k1^2 + b\, \k2^2 + 2\, c \,\k1 \cdot \k2 + 2\, d
\cdot\k1 + 2 \, e \cdot \k2 + f, 
\label{eq:sumaixi}
\end{equation}
where $a$, $b$, $c$, $d^\mu$, $e^\mu$ and $f$ are directly readable from the
actual graph: $a (b) = \sum x_i$, where the sum runs over the
legs in the $\k1$ $(\k2)$ loop, and $c = \sum x_i$ with the sum running over
the legs common to both loops.  Specifically, in the pentabox graph we have
\begin{eqnarray}
\label{eq:abc}
a &=& x_1+x_2+x_3+x_4+x_7\nonumber \\
b &=& x_5+x_6+x_7\nonumber \\
c &=& -x_7\nonumber \\
d^\mu &=& x_2 \,p_1^\mu + x_3 \,p_{12}^\mu + x_4\, p_{123}^\mu\nonumber \\
e^\mu &=& x_5 \, p_{123}^\mu\nonumber \\
f &=& x_3\, s,
\end{eqnarray}
where $p_{12} = p_1+p_2$ and $p_{123} = p_1+p_2+p_3$.  In all cases, the $a$,
$b$, $c$, $d^\mu$, $e^\mu$ and $f$ are linear in the $x_i$.

With the change of variables
\begin{eqnarray}
\label{eq:k1shift}
\k1^\mu &\to& \K1^\mu -\frac{c \K2^\mu}{a}
+\X^\mu, \\ 
\label{eq:k2shift}
\k2^\mu &\to& \K2^\mu + \Y^\mu,
\end{eqnarray}
where
\begin{equation}
\label{eq:XYdef}
\X^\mu = \frac{ce^\mu-bd^\mu}{\P}, \qquad \Y^\mu = \frac{cd^\mu-ae^\mu}{\P},
\end{equation}
and
\begin{equation}
\label{eq:P}
\P = ab-c^2,
\end{equation}
we can diagonalise Eq.~(\ref{eq:sumaixi}), so that
\begin{equation}
\sum_{i=1}^n x_iA_i
=
 a \K1^2+\frac{\P}{a}\K2^2 +\frac{\Q}{\P},
\end{equation}
with
\begin{equation}
\label{eq:Q}
\Q = -a\,e^2 - b\,d^2+2\,c\,e\cdot d+f \,\P.
\end{equation}
The scalar loop integral can be cast in the form
\begin{equation}
\Jd[1]
= \Dx
\Measure1 \Measure2 \;\exp\lq a \K1^2+\frac{\P}{a}\K2^2 +\frac{\Q}{\P}
\rq, 
\end{equation}
and the Gaussian integrals over the shifted loop momenta are 
evaluated to
\begin{equation}
\Jd[1]
= \Dx ~ \I,
\end{equation}
the integrand $\I$ being given by
\begin{equation}
\label{eq:def_I}
\I =
\frac{1}{\P^{D/2}}
~\exp\left (\frac{\Q}{\P} \right).
\end{equation}
Similarly, the tensor integrals can be easily obtained by using identities 
such as
\begin{eqnarray}
\Measure1 ~\K1^\mu ~\exp\left ( a \K1^2 \right) &=& 0,\\
\Measure1 ~\K1^\mu\K1^\nu ~\exp\left ( a \K1^2 \right) &=& 
-\frac{1}{2a}~{g^{\mu\nu}} ~\frac{1}{a^{D/2}},\\
\Measure1 ~\K1^\mu\K1^\nu\K1^\rho\K1^\sigma ~\exp\left ( a \K1^2 \right) &=& 
\frac{1}{4a^2} \left\{g^{\mu\nu}g^{\rho\sigma}
+g^{\mu\rho}g^{\nu\sigma}
+g^{\mu\sigma}g^{\nu\rho}\right\} \frac{1}{a^{D/2}} .
\end{eqnarray}
To give a concrete example, we consider the tensor integral associated with
$\k1^\mu$
\begin{eqnarray}
\label{eq:k1_mu}
\Jd[\k1^\mu] &=& 
\Dx
\nonumber \\
&\times &
\Measure1
\Measure2
~\left\{ \K1^\mu -\frac{c \K2^\mu}{a}+\X^\mu \right\}  
~\exp\left ( a \K1^2+\frac{\P}{a}\K2^2 +\frac{\Q}{\P} \right)\nonumber \\
&=&
 \Dx
~ \X^\mu \,\I.
\end{eqnarray}
Recalling the definition~(\ref{eq:XYdef}), we see that $\X^\mu$ consists of
the ratio of a set of bilinears in $x_i$ divided by $\P$.  We can therefore
absorb the factors of $x_i$ into ${\cal D}x$ (see Eq.~(\ref{eq:Dx})) by
increasing the power to which the $i$-th propagator is raised
\begin{equation}
\label{eq:times_x_i}
\frac{(-1)^{\nu_i} x_i^{\nu_i-1}}{\Gamma(\nu_i)} ~x_i \ \Longrightarrow \ 
-\nu_i \frac{(-1)^{\nu_i+1} x_i^{\nu_i}}{\Gamma(\nu_i+1)} \equiv 
-\nu_i {\bf  i^{\boldsymbol{+}}},
\end{equation}
while the factor $\P$ can be absorbed into $\I$ (see Eq.~(\ref{eq:def_I}))
\begin{equation}
\frac{1}{\P^{D/2}} ~\frac{1}{\P} \ \Longrightarrow\ 
\frac{1}{\P^{(D +2)/2}},
\end{equation}
so that $1/\P$ acts as a dimension increaser
\beq
\label{eq:defdpm}
\frac{1}{\P}  \ \Longrightarrow\ \dplus, \quad\quad\quad 
\P  \ \Longrightarrow\ \dminus,
\eeq
where
\beq
\label{eq:Pdm}
  {\bf d^{\boldsymbol{\pm}}} \Jd =  J^{D \pm 2}\(\{\nu_i\};\{Q_i^2\} \).
\eeq
In this way, each $x_i$ in the numerator increases by one the power of the
associated propagator, and each power of $\P$ in the denominator (numerator)
increases (decreases) the space-time dimension $D$ by two.
Schematically we have
\begin{equation}
\Jd[\k1^\mu] = \sum  \nu_i\nu_j ~ p_k^\mu~ J^{D+2} \( \lg \ldots ,\nu_{i+1},
\ldots, 
\nu_{j+1}, \ldots \rg ; \lg Q_i^2 \rg \)[1]  \, ,
\end{equation}
where the summation runs over the elements of $(ce^\mu - bd^\mu)$ which
fix the values of $i$, $j$ and $p_k$.

For the case of the pentabox where $a,\ldots, f$ are given in
Eq.~(\ref{eq:abc}), we have from Eq.~(\ref{eq:XYdef})
\beq
\X^\mu = \frac{-x_7 x_5 \, p_{123}^\mu -  \(x_5+x_6+x_7\)\( x_2 \,p_1^\mu +
x_3 \,p_{12}^\mu + x_4\, p_{123}^\mu\)}{\P}
\eeq
so that Eq.~(\ref{eq:k1_mu}) becomes
\beq
J^D [\k1^\mu] = \lq -\p5\p7 p_{123}^\mu -\(\p5+\p6+\p7\)\!\(\p2 p_1^\mu +\p3
p_{12}^\mu + \p4  p_{123}^\mu\) \rq \dplus J^{D},
\eeq
where we have used the shorthand notation $J^D=\Jd$.

By applying the integration-by-parts identities~(\ref{eq:IbyPa})
and~(\ref{eq:IbyPb}) we can now write this tensor integral directly in terms of
the set of simpler integrals.

For generic four-point integrals, we need tensor integrals with up to four
free indices, each associated with a Lorentz index of an external leg.
Integrals with higher powers of the loop momenta are of course possible, but
must yield dot products with other momenta when the available free Lorentz
indices are saturated.  In many cases, these dot products can be immediately
expressed in terms of the propagators and canceled through (see
Eq.~(\ref{eq:dot_prod})). 

Nevertheless, the procedure previously described can be iterated ad libitum
and we can express every tensor integrals in terms of scalar ones with
increased powers of the propagators and dimension $D$.
For example, we have
\begin{eqnarray}
J^D[\k2^\mu] &=& \Dx ~\Y^\mu ~\I,\\
J^D[\k1^\mu\k1^\nu] &=& \Dx~ \left(\X^\mu\X^\nu-\frac{b}{2P}g^{\mu\nu} \right)
~\I,\\
J^D[\k1^\mu\k2^\nu] &=& \Dx~ \left(\X^\mu\Y^\nu+\frac{c}{2P}g^{\mu\nu} \right)
~\I,\\
J^D[\k2^\mu\k2^\nu] &=& \Dx~ \left(\Y^\mu\Y^\nu-\frac{a}{2P}g^{\mu\nu} \right)
~\I,\\
J^D[\k1^\mu\k1^\nu\k1^\rho] &=& \Dx~ \left(\X^\mu\X^\nu\X^\rho-\frac{b}{2\P}
\lg g^{\mu\nu}\X^\rho + g^{\mu\rho}\X^\nu + g^{\nu\rho}\X^\mu \rg \right)
~\I,\\ 
J^D[\k1^\mu\k1^\nu\k2^\rho] &=& \Dx~ \left(\X^\mu\X^\nu\Y^\rho
-\frac{b}{2\P}g^{\mu\nu}\Y^\rho 
+\frac{c}{2\P}\{g^{\mu\rho}\X^\nu + g^{\nu\rho}\X^\mu \} 
\right) ~\I,\\
J^D[\k1^\mu\k2^\nu\k2^\rho] &=& \Dx~ \left(\X^\mu\Y^\nu\Y^\rho
-\frac{a}{2\P}g^{\nu\rho}\X^\mu 
+\frac{c}{2\P}\{g^{\mu\nu}\Y^\rho + g^{\mu\rho}\Y^\nu \} 
\right) ~\I,\\
J^D[\k2^\mu\k2^\nu\k2^\rho] &=& \Dx~ \left(\Y^\mu\Y^\nu\Y^\rho-\frac{a}{2\P}
\{g^{\mu\nu}\Y^\rho + g^{\mu\rho}\Y^\nu + g^{\nu\rho}\Y^\mu \} \right) ~\I,\\
J^D[\k1^\mu\k1^\nu\k1^\rho\k1^\sigma] &=& 
\Dx~ \Biggl(
\X^\mu\X^\nu\X^\rho\X^\sigma 
+\frac{b^2}{4\P^2}\{g^{\mu\nu}g^{\rho\sigma}+ g^{\mu\rho}g^{\nu\sigma}
+g^{\mu\sigma}g^{\nu\rho}\}
\nonumber \\
&&{} -\frac{b}{2\P}
\{g^{\mu\nu}\X^\rho\X^\sigma + g^{\mu \rho}\X^\nu\X^\sigma 
+ g^{\mu \sigma}\X^\nu\X^\rho + g^{\nu \rho}\X^\mu\X^\sigma \nonumber\\
&&{}
+ g^{\nu \sigma}\X^\mu\X^\rho + g^{\rho \sigma}\X^\mu\X^\nu\} \Biggr) ~\I, \\
J^D[\k1^\mu\k1^\nu\k1^\rho\k2^\sigma] &=& 
\Dx~ \Biggl(
\X^\mu\X^\nu\X^\rho\Y^\sigma
-\frac{b}{2\P}
\{g^{\mu\nu}\X^\rho 
+ g^{\mu\rho}\X^\nu 
+ g^{\nu\rho}\X^\mu\} \Y^\sigma\nonumber \\
&&{}
+\frac{c}{2\P}
\{g^{\mu\sigma}\X^\nu\X^\rho 
+ g^{\nu\sigma}\X^\mu\X^\rho 
+ g^{\rho\sigma}\X^\mu\X^\nu\} \nonumber \\
&&{}
-\frac{bc}{4\P^2}\{g^{\mu\nu}g^{\rho\sigma}+ g^{\mu\rho}g^{\nu\sigma}
+g^{\mu\sigma}g^{\nu\rho}\}
\Biggr) ~\I,\\
J^D[\k1^\mu\k1^\nu\k2^\rho\k2^\sigma] &=& 
\Dx~ \Biggl(
\X^\mu\X^\nu\Y^\rho\Y^\sigma
-\frac{a}{2\P}
g^{\rho\sigma}\X^\mu\X^\nu
-\frac{b}{2\P}
g^{\mu\nu}\Y^\rho\Y^\sigma
\nonumber \\
&&{}
+\frac{c}{2\P}
\{g^{\mu\rho}\X^\nu\Y^\sigma 
+ g^{\nu\rho}\X^\mu\Y^\sigma
+ g^{\mu\sigma}\X^\nu\Y^\rho 
+ g^{\nu\sigma}\X^\mu\Y^\rho  \} \nonumber \\
&&{}
+\frac{ab}{4\P^2}g^{\mu\nu}g^{\rho\sigma}
+\frac{c^2}{4\P^2}\{g^{\mu\rho}g^{\nu\sigma}+g^{\mu\sigma}g^{\nu\rho}\}
\Biggr) ~\I,\\
J^D[\k1^\mu\k2^\nu\k2^\rho\k2^\sigma] &=& 
\Dx~ \Biggl(
\X^\mu\Y^\nu\Y^\rho\Y^\sigma
-\frac{a}{2\P}
\{g^{\nu\rho}\Y^\sigma +g^{\nu\sigma}\Y^\rho
+g^{\rho\sigma}\Y^\nu \}\X^\mu \nonumber \\
&&{}
+\frac{c}{2\P}
\{g^{\mu\nu}\Y^\rho\Y^\sigma +g^{\mu\rho}\Y^\nu\Y^\sigma
+g^{\mu\sigma}\Y^\nu\Y^\rho  \} \nonumber \\
&&{}
-\frac{ac}{4\P^2}\{g^{\mu\nu}g^{\rho\sigma}+ g^{\mu\rho}g^{\nu\sigma}
+g^{\mu\sigma}g^{\nu\rho}\}
\Biggr) ~\I,\\
J^D[\k2^\mu\k2^\nu\k2^\rho\k2^\sigma] &=& 
\Dx~ \Biggl(
\Y^\mu\Y^\nu\Y^\rho\Y^\sigma
+\frac{a^2}{4\P^2}\{g^{\mu\nu}g^{\rho\sigma}+ g^{\mu\rho}g^{\nu\sigma}
+g^{\mu\sigma}g^{\nu\rho}\}\nonumber\\
&&{}
-\frac{a}{2\P}
\{g^{\mu\nu}\Y^\rho\Y^\sigma + g^{\mu \rho}\Y^\nu\Y^\sigma 
+ g^{\mu \sigma}\Y^\nu\Y^\rho + g^{\nu \rho}\Y^\mu\Y^\sigma \nonumber\\
&&{}
+ g^{\nu \sigma}\Y^\mu\Y^\rho + g^{\rho \sigma}\Y^\mu\Y^\nu\} \Biggr) ~\I.
\end{eqnarray}

Note that these expressions are valid for arbitrary two-loop integrals and to
use them all that needs to be done is to identify $a$, $b$, $c$, $d^\mu$,
$e^\mu$ and $f$, which can easily be read off from the graph, and to
construct $\X^\mu$ and $\Y^\mu$.  The powers of $x_i$ and $\P$ can then be
exchanged for scalar integrals with higher $\nu_i$ and higher $D$. This
procedure is straightforward to implement in an algebraic program.

Of course, rewriting the tensor integrals in this way is only useful if the
relevant scalar integrals in higher $D$ and with arbitrary powers of the
propagators are known.  For the pentabox, this is the case since repeated
application of the integration-by-parts identities~(\ref{eq:IbyPa})
and~(\ref{eq:IbyPb}) can always be used to reduce the integrals to one of the
master-topology integrals of Eqs.~(\ref{eq:masterA})--(\ref{eq:masterC}), for
which general results are given in the next section.

However, there is one further simplification of note.  When in the numerator
we have a dot product that can be rewritten in terms of the difference of
propagators, it is easiest to directly cancel off one of the propagator
powers. For example, if we have to deal with $\k1\cdot p_1$, it is more
efficient to make the replacement
\begin{equation}
\label{eq:dot_prod}
2 \,\k1\cdot p_1 = (\k1+p_1)^2-\k1^2 = A_2 - A_1 \equiv \m2 - \m1.
\end{equation}
The integration-by-parts identities~(\ref{eq:IbyPa}) and~(\ref{eq:IbyPb}) can
still be used to further reduce the powers of $\nu_1$, $\nu_4$, $\nu_5$ and
$\nu_6$.

\begin{figure}[t]
\begin{center}
\epsfig{file=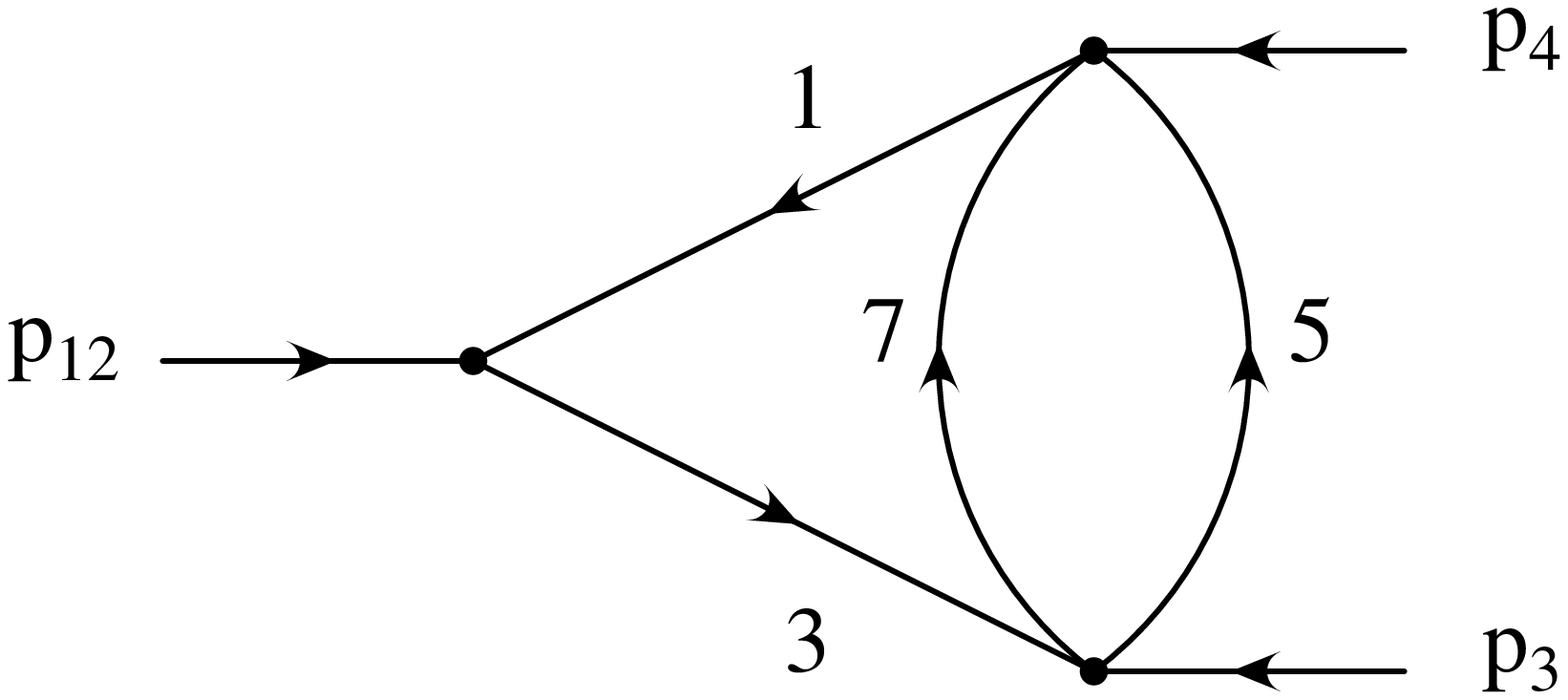,height=6cm}
\end{center}
\ccaption{}{\label{fig:E}
One-loop self-energy insertion into a one-loop triangle.  The propagators are
 labelled according to Eq.~(\ref{eq:props}) and are each raised to the
 $\nu_i$ power.}
\end{figure}

Following this procedure, we can pinch completely the second propagator, so
that we immediately obtain self-energy insertions to one-loop triangle graphs
of the type shown in Fig.~\ref{fig:E}.  These can be written directly in
terms of $\Gamma$ functions
\begin{eqnarray}
J^D(\nu_1, 0,\nu_3,0,\nu_5,0,\nu_7;s,t) &=&\BUB(\nu_5,\nu_7)
~I^D_3\(\nu_5+\nu_7-\halfD,\nu_1,\nu_3;s\),\\
J^D(0, 0,\nu_3,\nu_4,0,\nu_6,\nu_7;s,t) &=&
\BUB(\nu_6,\nu_7)~I^D_3\(\nu_4,\nu_6+\nu_7-\halfD,\nu_3;s\),\\
J^D(0, 0,\nu_3,0,\nu_5,\nu_6,\nu_7;s,t) &=& 
\BUB(\nu_3,\nu_7)~I^D_3\(\nu_5,\nu_3+\nu_7-\halfD,\nu_6;s\),
\end{eqnarray}  
where the self-energy factor is given by
\begin{equation}
\label{eq:bubble}
\BUB(\nu_1,\nu_2) = \pow
\frac{\G{\nu_1+\nu_2-\halfD} 
      \G{\halfD-\nu_1}\G{\halfD-\nu_2}}
{\G{\nu_1}\G{\nu_2}\G{D-\nu_1-\nu_2}},
\end{equation}
and the one-loop triangle integral with arbitrary powers and 
with one external mass scale $Q^2$ is given by (see for example~\cite{AGO1})
\begin{equation}
I_3^D\(\nu_1,\nu_2,\nu_3;Q^2\) =
 \pow \(Q^2\)^{\halfD-\n{123}}
\frac{\G{\halfD-\n{12}}\G{\halfD-\n{13}}\G{\n{123}-\halfD}}
{\G{\nu_2}\G{\nu_3}\G{D-\n{123}}},
\label{eq:onelooptri}
\end{equation}
where we have introduced the shorthand $\nu_{ij} = \nu_i+\nu_j$, $\nu_{ijk} =
\nu_i+\nu_j+\nu_k$, etc.  
  
\section{The master topologies}
\label{sec:master}

In this section, we collect the explicit expressions in terms of
hypergeometric functions for the general master-topology integrals $A^D$
($B^D$) and $C^D$, with arbitrary powers of propagators and general dimension
$D$.  We also show how to relate these to simpler pinched integrals and we
compute explicitly the integrals that we need to evaluate all the tensor
integrals in a closed analytical form and as an expansion in $\ep$
($D=4-2\ep$).

\subsection{Topology $\boldsymbol{A}$}

\begin{figure}[t]
\begin{center}
\epsfig{file=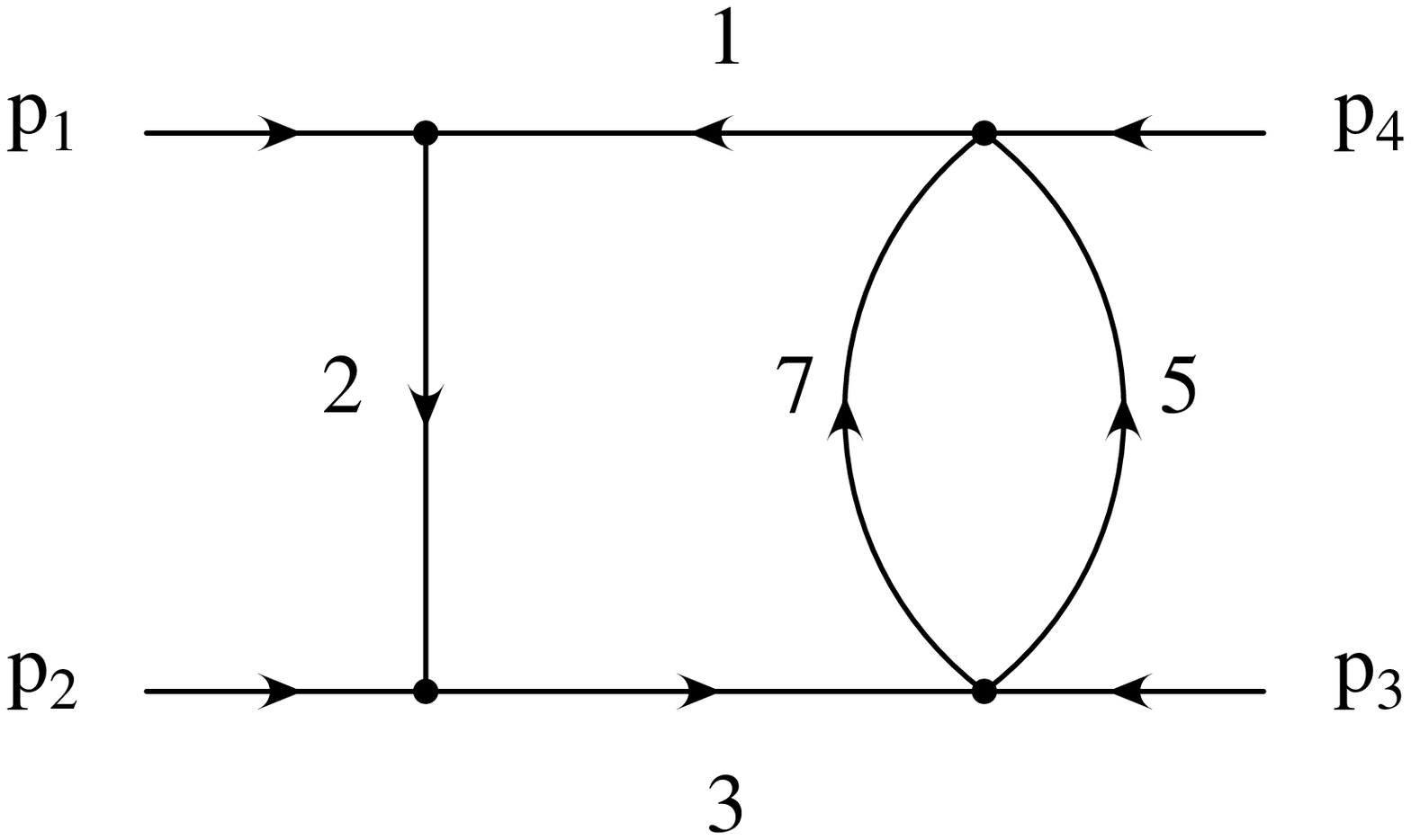,height=5cm}
\end{center}
\ccaption{}{\label{fig:A}
The master topology $A$.  The propagators are labelled according to
Eq.~(\ref{eq:props}) and are each raised to the $\nu_i$ power.}
\end{figure}

The master-topology integral $\Abox$ is shown in Fig.~\ref{fig:A} and, with
respect to the propagator factors of Eq.~(\ref{eq:props}), is defined as
\begin{equation}
\Abox = \int  \frac{d^D\k1}{i\pi^{D/2}}
\int \frac{d^D\k2}{i\pi^{D/2}} \frac{1}{
A_1^{\nu_1}A_2^{\nu_2}A_3^{\nu_3}A_5^{\nu_5}A_7^{\nu_7}}.
\label{eq:Adef}
\end{equation}
This integral is related by a factor to the ordinary one-loop box integral
\begin{equation}
\Abox = 
\BUB(\nu_5,\nu_7)\,I_4^D\left(\nu_1,\nu_2,\nu_3,\nu_5+\nu_7-\halfD;s,t\right),
\end{equation}
where $\BUB(\nu_5,\nu_7)$ is given in Eq.~(\ref{eq:bubble}) and
\begin{equation}
I_4^D(\nu_1,\nu_2,\nu_3,\nu_4;s,t) = \int \frac{d^D\k1}{i\pi^{D/2}}
\frac{1}{A_1^{\nu_1}A_2^{\nu_2}A_3^{\nu_3}A_4^{\nu_4}}.
\label{eq:boxdef}
\end{equation}
Expressions for the one-loop box integral with general powers of the
propagators are easily obtained~\cite{AGO2,AGO1,VanNeerven}.  In the kinematic
region $|t| < |s|$ we find~\cite{AGO2}
\begin{eqnarray}
\label{eq:oneloopbox}
\lefteqn{\hspace{-1.3cm}
I_4^D(\nu_1,\nu_2,\nu_3,\nu_4;s,t) = \pow \; s^{\halfD-\n{1234}} \;}
\nonumber \\
\times & \Biggl[ &
\frac{\G{\n{1234}-\halfD}\G{\halfD-\n{234}}\G{\halfD-\n{124}}}{\G{\n{1}}
\G{\n{3}}\G{D-\n{1234}}} 
\nonumber \\
&& \times ~\ftt \left( \n{2}, \n{4},\n{1234}-\halfD,
1+\n{234}-\halfD,1+\n{124}-\halfD,-\frac{t}{s}\right)
\nonumber \\
&+& 
\frac{\G{\n{234}-\halfD}\G{\halfD-\n{34}}\G{\n{3}-\n{1}}\G{\halfD-\n{23}}}{\G{\n{2}}\G{\n{3}}\G{\n{4}}
\G{D-\n{1234}}}
\nonumber \\
&& \times \left(\frac{t}{s}\right)^{\halfD-\n{234}}
\ftt \left( \n{1}, \halfD-\n{34},\halfD-\n{23}, 1+\n{1}-\n{3}, 1+\halfD-\n{234},-\frac{t}{s}\right)
\nonumber \\
&+& 
\frac{\G{\n{124}-\halfD}\G{\halfD-\n{14}}\G{\n{1}-\n{3}}\G{\halfD-\n{12}}}{\G{\n{1}}\G{\n{2}}\G{\n{4}}
\G{D-\n{1234}}}
\nonumber \\
&&  \times \left(\frac{t}{s}\right)^{\halfD-\n{124}}
\ftt \left( \n{3}, \halfD-\n{14},\halfD-\n{12}, 1+\n{3}-\n{1},
1+\halfD-\n{124},-\frac{t}{s}\)
\Biggr].
\end{eqnarray}
Expressions for $I^D_4$ valid for $|s| < |t|$ can
be obtained by the exchanges $\nu_1 \leftrightarrow \nu_2$, $\nu_3
\leftrightarrow \nu_4$ and $s \leftrightarrow t$ in
Eq.~(\ref{eq:oneloopbox}).

Although these integrals appear with a variety of integer values for $\nu_1$,
$\nu_2$ and $\nu_3$ in dimensions higher than $D=4-2\epsilon$, we can
systematically relate them to simpler integrals by the integration-by-parts
identities.  In fact, we find
\begin{eqnarray}
\label{eq:boxred1}
s \,\p1  I_4^{D} \!\!&=&\!\! -\(D-\nu_{12334}\)\, I_4^{D}
+\(\p1+\p2+\p4\)\m3 I_4^D,
\\
\(D-2-\nu_{1223}\) \,\p2 I_4^{D} \!\!&=&\!\! 
\(D-2-\nu_{1344}\) \,\p4 I_4^{D} 
+\(\nu_{2}-\nu_{4}\)\(\p1+\p3\) I_4^D,
\\  \label{eq:boxred3}
s \,\p3 I_4^{D} \!\!&=&\!\! -\(D-\nu_{11234}\) \, I_4^{D} 
+\(\p2+\p3+\p4\)\m1 I_4^D,\hspace{2cm}
\end{eqnarray}
where we have used the shorthand notation $I_4^D =
I_4^{D}(\nu_1,\nu_2,\nu_3,\nu_4;s,t)$ and 
$\nu_{ijjk} = \nu_{i}+2\nu_{j}+\nu_{k}$, etc.

Repeated application of these identities reduces $\nu_1$, $\nu_2$ and $\nu_3$
to unity together with integrals where one of the propagators is pinched out,
yielding a one-loop triangle (see Eq.~(\ref{eq:onelooptri}))
\begin{eqnarray}
I_4^D\(\nu_1,\nu_2,0,\nu_4;s,t\) = I_3^D\(\nu_1,\nu_2,\nu_4;t\),\\
I_4^D\(0,\nu_2,\nu_3,\nu_4;s,t\) = I_3^D\(\nu_3,\nu_2,\nu_4;t\).
\end{eqnarray}
Subsequent application of
\begin{eqnarray}
\label{eq:boxred4}
t\, \p4 I_4^{D} &=& -\(D-\nu_{12234}\) \, I_4^{D} +
\(\p1+\p3+\p4\)\m2 I_4^D, 
\end{eqnarray}
can be used to control the power of $\nu_4$ and form the pinched triangle
integral
\begin{equation}
I_4^D(\nu_1,0,\nu_3,\nu_4;s,t) = I_3^D(\nu_4,\nu_1,\nu_3;s).
\end{equation}
Equation~(\ref{eq:boxred4}) should be used until $\nu_4 =
2-D/2$, corresponding to $\nu_5=\nu_7 = 1$,  that is the
master integral for the $A$ topology.

The dimensional-shift identity for the master integral is obtained 
applying
\beq
\dminus = \lq (\p3+\p2+\p1)(\p5+\p7) +\p5\p7\rq 
\eeq
to $A^{D+2}\(1,1,1,1,1;s,t\)$, and reducing the right-hand side with the help
of Eqs.~(\ref{eq:boxred1})--(\ref{eq:boxred3}) and~(\ref{eq:boxred4}):
\beqn
A^{D+2}\(1,1,1,1,1;s,t\) &=&
\frac{(D-4) s t^2}{3 (D-1) (3 D-10) (3 D-8) (t+s)}  \, A^{D}\(1,1,1,1,1;s,t\) 
\nonumber\\
&+&   \frac{s \lq (D -4 )t+ (2 D -6) s \rq}
	{3 (D-2) (D-1) (3 D-8) (t+s)} \, I_3^D\(2-\halfD,1,1;s\) 
~\BUB(1,1)
\nonumber\\
&+& \frac{t}{3 (D-4) (D-1) (t+s)} \, I_2^D\(2-\halfD,1;t\) ~\BUB(1,1),
\eeqn
where
\beq
I_2^D\(\n{1},\n{2};Q^2\) = (Q^2)^{\halfD-\nu_{12}} ~\BUB(\nu_1,\nu_2).
\eeq
The master integral for the $A$ topology has all propagators raised to unit
powers and may written in terms of Gaussian hypergeometric
functions as~\cite{AGO2}
\begin{eqnarray}
A^D\(1,1,1,1,1; s,t\)
\!\!&=& \!\!-
\frac{\Gamma^2\left(2-\halfD\right)\Gamma^2\left(\halfD-1\right)\G{4-D}\G{D-3}}
{\G{\frac{3D}{2}-5} \G{D-2}\G{3-\halfD}}\nonumber \\
&\times &\Biggl[
(-t)^{D-5}~
\frac{\G{\halfD-2}}{
\G{2-\halfD}}
~\f21 \left(1,1,\halfD-1,\frac{s+t}{t}\right) \nonumber \\
&& {}+ (-s)^{D-5} ~ \G{D-3}
~\f21 \left(1,2-\halfD,\halfD-1,\frac{s+t}{s}\right) \Biggr].
\label{eq:nomass_nu4}
\end{eqnarray}
Making the expansion  in $\ep=D/2-2$ of the hypergeometric functions, we
obtain 
\beq
A^D\(1,1,1,1,1;s,t\)  
= \frac{\Gamma^3\(1-\ep\)\G{1+2 \ep}}{2 \, s \, \(1-2 \ep \) \,  \ep^3 \, 
\G{1-3\ep}}  \lq \(-s \)^{-2 \ep} A_1\(s, t\) +    
\(-t\)^{-2 \ep} A_2\(s, t\) 
\rq
\end{equation}
where $A_1\(s, t\) $ and $A_2\(s, t\) $ are given respectively by:
\begin{enumerate}

\item[1)] 
in the physical region $s>0$, $t<0 $:
\beqn
A_1\(s, t\) &= & \(-\frac{t}{s}\)^{-\ep}  \lg 
1 - \ep^2 \lq \li{\fracmut} - 2 \zeta_2 \rq 
- \ep^3 \lq \lit{\fracmut} + \sot{\fracmut} 
\right. \right. 
\nonumber \\ 
&& \left. 
{} -2 \zeta_3 \bigg] - \ep^4  \lq 
\lif{\fracmut}  + \st2{\fracmut} + \so3{\fracmut}
\right. \right. 
\nonumber \\ 
&& \left. \left.
{}+ 2 \zeta_2 \li{\fracmut} -
9 \zeta_4 \rq
\rg  +  {\cal O} \( \ep^{5} \), 
\\
A_2\(s, t\) &= & 1 + \ep \log \(-\frac{t}{s}\) - \ep^2 \li{\fracmus} 
- \ep^3 \lit{\fracmus} - \ep^4 \lif{\fracmus} 
\nonumber 
\\ &&{}  +  {\cal O} \( \ep^{5} \), 
\end{eqnarray}
\item[2)] while in the region $s<0$, $t<0$:
\begin{eqnarray}
A_1\(s, t\) & = & \(\frac{t}{s}\)^{-\ep} \lg 1+ \ep^2 \lq \li{\fracmtu} +
  \frac{1}{2} \log^2 \(\fracmut \) - \frac{\pi^2}{2} \rq \right. 
 - \ep^3 \lq 2\lit{\fracmtu}\right. 
\nonumber \\ 
&& \left.\left. {}- \sot{\fracmtu} -\zeta_3 
+ \log \(\fracmut\) \( \li{\fracmtu} +  \frac{5 }{6}\pi^2\) \rq \right. 
\nonumber \\ 
&&  \left. {} + \ep^4 \lq 
\so3{\fracmtu} -2 \st2{\fracmtu} + 4 \lif{\fracmtu} -\frac{23}{180}\pi^4
\right. \right. 
\nonumber \\ 
&& \left. \left. {}+ \frac{1}{24}  
 \log^4 \(\fracmut\) + \log \( \fracmut\) \( 2 \lit{\fracmtu} 
-\sot{\fracmtu} -\zeta_3  \) \rq  \right. 
\nonumber \\ && 
\left. {} + \frac{\pi^2}{3} \li{\fracmtu} - 
\frac{\pi^2}{4} \log^2\(\fracmut \) 
+ \frac{1}{2}\log^2 \(\fracmut \) \li{\fracmtu} \rg \nonumber 
\\ && {} + {\cal O} \( \ep^{5} \), 
\\
A_2\(s, t\) &= & 1+ \ep \log \(\frac{t}{s} \) 
+ \ep^2 \lq \li{\fracmsu} + \frac{1}{2} \log^2 \( \fracmus\)  
- \frac{\pi^2}{3} \rq 
\nonumber \\
&& {} - \ep^3 \lq \lit{\fracmsu} +   \frac{\pi^2}{3} \log \(\fracmus \) 
-\frac{1}{6} \log^3 \(\fracmus \) \rq 
\nonumber \\
&& {} + \ep^4 \lq \lif{\fracmsu} - \frac{\pi^4}{45} - 
\frac{\pi^2}{6} \log^2 \(\fracmus \) + \frac{1}{24}
  \log^4 \( \fracmus\) \rq   \nonumber \\
&& {}+  {\cal O} \( \ep^{5} \).
\end{eqnarray}
\end{enumerate}


\subsection{Topology $\boldsymbol{C}$}

\begin{figure}[t]
\begin{center}
\epsfig{file=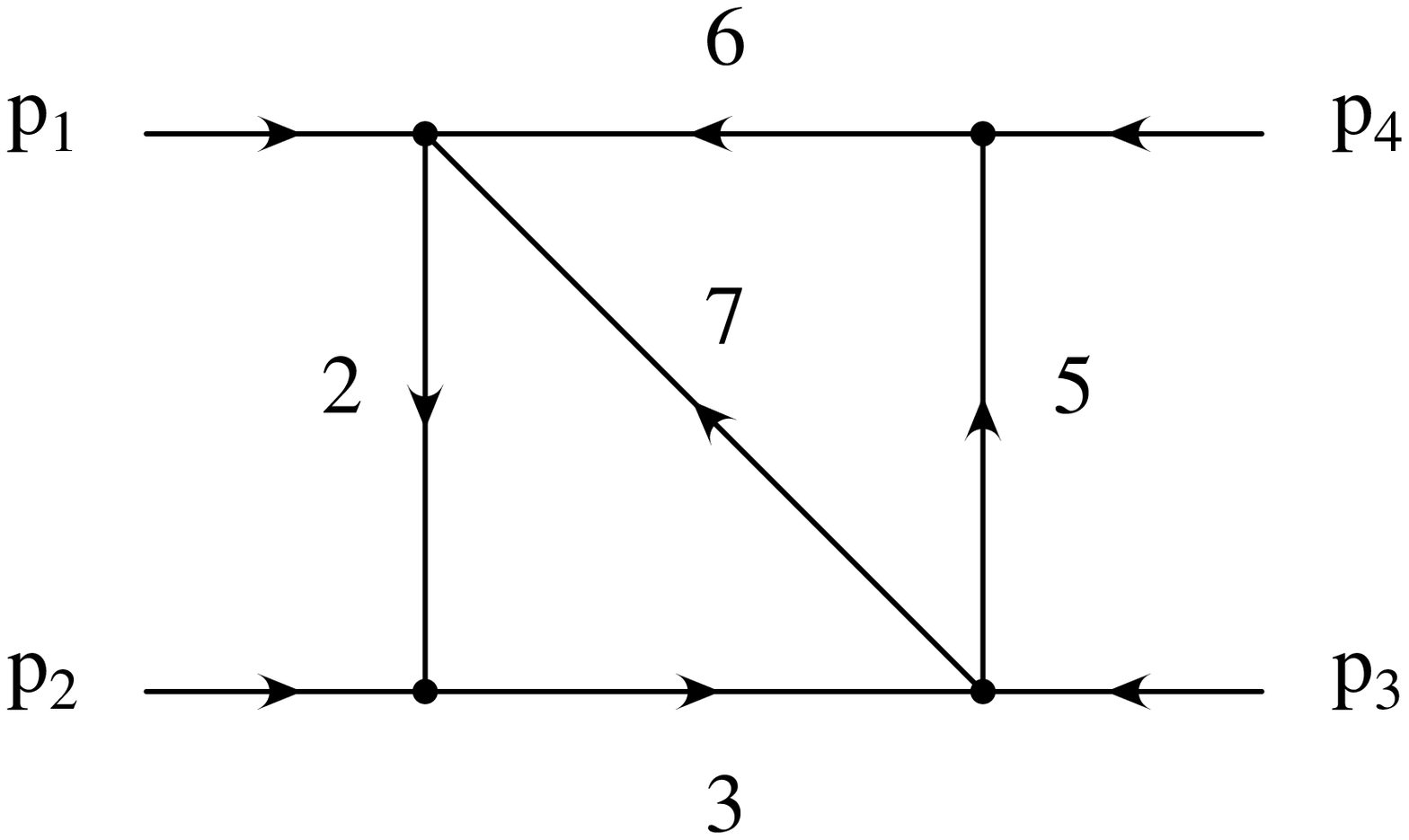,height=5cm}
\end{center}
\ccaption{}{\label{fig:C}
The master topology $C$.  The propagators are labelled according to
Eq.~(\ref{eq:props}) and are each raised to the $\nu_i$ power.}
\end{figure}

The diagonal-box master-topology integral is shown in Fig.~\ref{fig:C} and with
respect to the propagator factors of Eq.~(\ref{eq:props}) is defined as
\begin{equation}
\Cbox
=
\int 
\frac{d^D\k1}{i\pi^{D/2}}
\int
\frac{d^D\k2}{i\pi^{D/2}}
\frac{1}{
A_2^{\nu_2}A_3^{\nu_3}A_5^{\nu_5}A_6^{\nu_6}A_7^{\nu_7}}.
\label{eq:Cdef}
\end{equation}
For this graph, it is easy to read off the values of $\Q$ and $\P$ of
Eqs.~(\ref{eq:P}) and~(\ref{eq:Q})
\begin{eqnarray}
\Q &=& x_3 x_6 x_7 \, s + x_2 x_5 x_7 \,t,\\
\label{eq:P_diag_box}
\P &=& x_7(x_2+x_3+x_5+x_6) + (x_2+x_3)(x_5+x_6).
\end{eqnarray}
The integral can be written as a single Mellin-Barnes integral (see for
example Refs.~\cite{Smirnov2,Bas})
\begin{eqnarray}
\lefteqn{\Cbox = 
\frac{(-1)^D \; s^{D-\n{23567}}}
{\G{\n{2}} \G{\n{3}}\G{\n{5}} \G{\n{6}} \G{\n{7}} }}\nonumber \\
&&{}\times \frac{\G{\halfD -\n{7}}  \G{\halfD-\n{56}} 
\G{\halfD-\n{23}}}{\G{D-\n{237}} \G{D-\n{567}} 
\G{\frac{3}{2}D-\n{23567}} } \, \int^{i \infty}_{-i\infty} \frac{d\al}{2\pi i}
\G{-\al}    \G{D+\n{2}-\n{23567}-\al} 
\nonumber \\
&&{}\times  \G{D+\n{5}-\n{23567}-\al} \G{\n{23567}-D+\al} \G{\n{3}+\al}
\G{\n{6}+\al} \( \frac{t}{s}\)^\al,
\end{eqnarray}
where the path of integration over $\al$ must be chosen so that to separate
the poles coming from  $\G{\ldots-\al}$  from those coming from
$\G{\ldots+\al}$.  
In the kinematic region $|t| < |s|$ the  contour at the infinity must be
closed to the right, and the expression we obtain is
\begin{eqnarray}
\label{eq:Cres}
\lefteqn{\hspace{-1cm}
\Cbox = (-1)^D \; s^{D-\n{23567}} \;
\frac{\G{\halfD-\n{23}}\G{\halfD-\n{56}}\G{\halfD-\n{7}}}
{\G{\n{7}}\G{\frac{3}{2}D-\n{23567}}}\nonumber }
\\  \times &\Biggr[& 
\frac{\G{\n{23567}-D} \G{D- \n{2367}}    \G{D-\n{3567}}}
{\G{\n{2}} \G{\n{5}}\G{D-\n{567}} \G{D-\n{237}}} 
\nonumber \\ && 
\times \,\ftt\left( \n{3},\n{6}, \n{23567}-D, 1-D+\n{2367},1-D+\n{3567},
-\frac{t}{s}\right) 
\nonumber \\ &+& 
\frac{\G{\n{2}-\n{5}} \G{D- \n{267}}   \G{\n{2367}-D}}
{\G{\n{2}} \G{\n{3}}  \G{\n{6}} \G{D-\n{567}}} 
\nonumber \\ && 
\times  \left(\frac{t}{s}\right)^{D-\n{2367}} \;\ftt \left(\n{5}, D-\n{267},
D-\n{237}, 1+D-\n{2367}, 1+\n{5}-\n{2},\, -\frac{t}{s}\right)
\nonumber \\ &+& 
\frac{\G{\n{5}-\n{2}} \G{D- \n{357}}   \G{\n{3567}-D} }
{\G{\n{3}} \G{\n{5}}  \G{\n{6}} \G{D-\n{237}}} 
\nonumber \\ && 
\times \left(\frac{t}{s}\right)^{D-\n{3567}} \;
\ftt \left( \n{2},D-\n{357}, D-\n{567}, 1+D-\n{3567},1+\n{2}-\n{5},
-\frac{t}{s}\right)  \Biggl].\nonumber \\
\end{eqnarray}
The solution valid when $|s| < |t|$ can be obtained from Eq.~(\ref{eq:Cres})
by the exchanges
\begin{equation}
s \leftrightarrow t, \qquad 
\n{2} \leftrightarrow \n{3}, \qquad 
\n{5} \leftrightarrow \n{6}.
\end{equation}
The expression for the diagonal box (\ref{eq:Cres}) has an apparent
singularity when $\n{2}-\n{5}$ is an integer which cancels in the actual
evaluation of the diagram.

In addition, although in principle we need to know this integral for
arbitrary powers and arbitrary dimension, further simplifications occur.
Integration by parts yields the following relations ($C^D=\Cbox$)
\beqn
\(D-2-2\nu_{23}\) \p2 C^D &=& \(D-2-2\nu_7\) \p7 C^D
-\(D-2-2\nu_{23}\)\p3 C^D,\\
\(D-2-2\nu_{56}\) \p6 C^D &=& \(D-2-2\nu_7\) \p7 C^D
-\(D-2-2\nu_{56}\) \p5 C^D,
\end{eqnarray}
so that we can reduce both $\nu_2$ and $\nu_6$ to
unity at the expense of increasing $\nu_3$ and $\nu_5$ together with $\nu_7$.
We now reduce $\nu_3$ and $\nu_5$ to unity using the relations
\beqn
s\(D-2-2\nu_{23}\) \p3 C^D &=& -\(D-1-\nu_{237}\)\(3D-2\nu_{235667}\) C^D
\\
&&{} +2\(D-1-\nu_{237}\) \p5\m6 C^D+ \(D-2-2\nu_{7}\) \p7 \m6 C^D,\nonumber \\
t\(D-2-2\nu_{56}\) \p5 C^D &=& -\(D-1-\nu_{567}\)\(3D-2\nu_{223567}\) C^D
\\
&& {}+2\(D-1-\nu_{567}\) \p3\m2 C^D+ \(D-2-2\nu_{7}\) \p7 \m2 C^D,\nonumber
\end{eqnarray}
which, because $\nu_2$ and $\nu_6$ are already unity, produces simpler
pinched integrals of the form
\begin{eqnarray}
C^{D}(0,\nu_3,\nu_5,\nu_6,\nu_7;s,t)= \Pi^D\(\nu_3,\nu_7\)\;
I^D_3\(\nu_5,\nu_6,\nu_3+\nu_7-\halfD;s\),
\nonumber \\
C^{D}(\nu_2,\nu_3,\nu_5,0,\nu_7;s,t) = \Pi^D\(\nu_5,\nu_7\) \;
I^D_3\(\nu_3,\nu_2,\nu_5+\nu_7-\halfD;t\),
\end{eqnarray}
where the self-energy factor $\Pi^D$ is defined in Eq.~(\ref{eq:bubble}) and
the one-loop triangle $I^D_3$ is given by Eq.~(\ref{eq:onelooptri}). 

When the outer propagators have unit powers, we can reduce $\nu_7$ using
\beqn
st\(D-2-2\nu_{7}\) \p7 C^D &=& -(s+t)\(D-3-\nu_7\)\(3D-10-2\nu_7\) C^D 
\\
&& {}+2 \(D-3-\nu_7\) \( t \ps5\m6 + s \ps6\m5 \) C^D \nonumber \\
&& {}+\(D-2-2\nu_7\) \( t \p7\m6 + s \p7 \m5 \) C^D.\nonumber
\end{eqnarray}
This equation is only valid when $\nu_2 = \nu_3 = \nu_5 = \nu_6 =
1$.
 
Finally, applying
\beq
\dminus = \p7\(\p2+\p3+\p5+\p6\) + \(\p2+\p3\)\(\p5+\p6\)
\eeq   
to $C^{D+2}\(1,1,1,1,1;s,t\)$, we can obtain the dimensional-shift equation
\beqn
C^{D+2}\(1,1,1,1,1;s,t\) &=& 
\frac{(D-4)^2 s^2 t^2} {3 (D-3) (D-2) (3 D-10) (3 D-8)(t+s)^2} \,
   C^{D}\(1,1,1,1,1;s,t\)
\nonumber\\
&&{} 	
   + \frac{s \lq (2 D -5) t+ (D -3) s \rq}{3 (D-3) (D-2)^2 (t+s)^2} \, 
   I_2^D\(2-\halfD,1;s\) \,\BUB\(1,1\)
\nonumber\\
&&{} 
	+ \frac{t \lq (D -3) t+(2 D -5) s \rq}{3 (D-3) (D-2)^2 (t+s)^2}\,
   I_2^D\(2-\halfD,1;t\)\,\BUB\(1,1\).
\eeqn
To summarize, we actually only need to know $C^D(1,1,1,1,1;s,t)$ rather
than the more general case.  With the propagator powers equal to unity, all
of the $\ftt$ functions of Eq.~(\ref{eq:Cres}) reduce to $\f21$.  To deal
with the pole in $(\nu_2-\nu_5)$ we proceed as in Ref.~\cite{AGO1}: we set
$\nu_2=\nu_5+\delta$, and, after performing an appropriate analytical
continuation, we take the limit $\delta \to 0 $.  
The final expression is given by
\begin{eqnarray}
\label{eq:Cexp}
C^D(1,1,1,1,1;s,t)&=&  -
\frac{\G{\halfD-1} \G{3-D}  \Gamma^2\(\halfD-2\) }
{  \G{\frac{3}{2} D -5}} \nonumber \\
&\times& \Biggr[ (-t)^{D-5}
\; \f21 \left(1,1,  D-2 , \frac{s+t}{t} \right)
\;  \nonumber \\
&&{} 
+  (-s)^{D-5}
\; \f21 \left(1,  1, D-2 , \frac{s+t}{s} \right)
\Biggl].
\end{eqnarray}
If we make a series expansion in $\ep$, we obtain
\beq
C^D\(1,1,1,1,1;s,t\) = 
\frac{\Gamma^3\(1-\ep\) \G{1+2 \ep}}{2 \(s+t\) \G{1-3\ep} \, \ep^3} 
\lq
\(-s \)^{-2 \ep} C\(s, t\) +    \(-t\)^{-2 \ep} C\(t, s\) 
\rq,
\eeq
where $C\(s, t\)$ is given respectively by: 
\begin{enumerate}
\item[1)]
in the physical region $s>0$, $t<0 $:
\beqn
C\(s,t\) &=& \log\(-\frac{t}{s}\) + 2 \ep \li{\fracmut} + 
 4 \ep^2 \lit{ \fracmut}  + 8 \ep^3 \lif{ \fracmut} 
\nonumber \\
&& {}+  {\cal O} \( \ep^{4} \),
\eeqn
\item[2)]
while in the region $s<0$, $t<0$:
\beqn
C\(s,t\) &=& \log \(\frac{t}{s} \) -2 \ep 
\lq \li{\fracmtu} +\frac{1}{2} \log^2 \( \fracmut \) - \frac{\pi^2}{3} \rq
\nonumber \\
&& {}
+ 4 \ep^2 \lq \lit{\fracmtu} -\frac{1}{6} \log^3 \( \fracmut \)  + 
\frac{\pi^2}{3} \log \( \fracmut \)\rq  
\nonumber \\
&&{}
-8 \ep^3 \lq \lif{\fracmtu} + \frac{1}{24} \log^4\(\fracmut \) -
\frac{\pi^2}{6} \log^2 \( \fracmut\)  - \frac{\pi^4}{45} \rq 
\nonumber \\
&& {}+  {\cal O} \( \ep^{4} \).
\eeqn
\end{enumerate}
Note that the prefactor of Eq.~(\ref{eq:Cexp}) indicates that the integral
diverges as $1/\epsilon^3$.  However,  the hypergeometric
functions conspire to remove the leading divergence and we reproduce the
result for $C^D(1,1,1,1,1;s,t)$ quoted in Ref.~\cite{Smirnov2}.

\section{Conclusions}
\label{sec:conc}

Finally we summarize what we have accomplished in this paper.  The pentabox
with light-like external legs is one of the two-loop box graphs needed for
the evaluation of Feynman diagrams for physical $2~\to~2$ scattering
processes, the others being the recently evaluated planar double-box
graph~\cite{Smirnov,Smirnov2} and non-planar double-box graph~\cite{Bas} as
well as one-loop box integrals with bubble insertions on one of the
propagators~\cite{AGO2}. Using integration by parts identities, we have
derived explicit analytic expressions for the scalar pentabox graph
$J^D\(1,1,1,1,1,1,1;s,t\)$ in terms of two simpler box integrals ($A$ and
$C$) with fewer propagators and given an explicit expansion in $\ep$ in terms
of polylogarithm functions (see Eq.~(\ref{eq:pentaboxts})).

In Section~\ref{sec:tensor} we demonstrated how generic two-loop integrals
could be expressed directly in terms of integrals in higher dimension with
shifted propagator powers. Similar results have been given by
Tarasov~\cite{Tar} using differential operators; however we believe our
method based on completing the square and Gaussian integration to be more
direct and simpler to apply to generic integrals where the quantities $a$,
$b$, $c$, $d^\mu$, $e^\mu$ and $f$ (see Eq.~(\ref{eq:sumaixi})) can easily be
read off from the graph and used to construct the vectors $\X^\mu$ and
$\Y^\nu$ (defined in Eq.~(\ref{eq:XYdef})), which therefore are linear in both
the external momenta and the $x_i$.  As detailed in Section~\ref{sec:tensor},
the powers of $x_i$ and $\P$ present in Eqs.~(4.22) and~(4.25)--(4.37) can
then be straightforwardly exchanged for scalar integrals with higher $\nu_i$
and higher $D$.

In the case of the pentabox, the tensor integrals can also be written in
terms of the same master-topology integrals as the scalar pentabox, but with
different powers of propagators and in higher dimension.  We give quite
general analytic formulae for the master-topology integrals with arbitrary
dimension and propagators.  For the $A$ topology we quote the results of
\cite{AGO2} while for the $C$ topology we employ the Mellin-Barnes integral
representation.  Other methods would also suffice.  In fact, it is not
necessary to know the master integrals in their full generality since further
integration-by-parts identities relate the various integrals.  Explicit
expressions for topology $A$ and $C$ are given in Eqs.~(\ref{eq:nomass_nu4})
and (\ref{eq:Cexp}) respectively, together with instructions on how to apply
them to more general integrals.

Finally, we note that the recurrence relations necessary for the complete
reduction of tensor integrals for the planar-box graph were described in
Ref.~\cite{Smirnov2}.  However, the situation for the non-planar box is less
well developed.  At present only one master integral is known, and the
solution of the recurrence relations for that topology is an open problem.

\section*{Acknowledgements}

We thank J.V.~Armitage, J.B.~Tausk, M.E.~Tejeda-Yeomans, J.J.~van der Bij and
M.~Zimmer for assistance and useful suggestions and D.~Broadhurst for his
encouragement. C.A. acknowledges the financial support of the Greek
Government and C.O. acknowledges the financial support of the INFN.
We gratefully acknowledge the support of the British Council and German
Academic Exchange Service under ARC project 1050.

\appendix
\section{Polylogarithms}
\label{sec:app}

The purpose of this appendix is to define the generalised polylogarithms that
occur in the expansion in $\ep$ of the pentabox scalar and tensor loop
integrals and to give useful identities amongst the polylogarithms.  In
Section~\ref{subsec:defs} we give the definitions of the polylogarithm
functions $\ss{n}{p}{x}$.  These functions are real when $x \le 1$ but they
develop an imaginary part for $x > 1$.  Analytic continuation formulae are
given in Section~\ref{subsec:anal}.  Finally, useful identities between
polylogarithms are listed in Section~\ref{subsec:ident}.

\subsection{Definition}
\label{subsec:defs}

The generalised polylogarithms of Nielsen are defined by
\beq
\ss{n}{p}{x} = \frac{(-1)^{n+p-1}}{(n-1)!\, p!} \int_0^1 dt \, 
  \frac{\log^{n-1}(t) \log^{p}(1-xt)}{t}, \quad \quad \quad n,p \ge 1, \ \
  x\le 1.
\eeq
For $p=1$ we find the usual polylogarithms
\beq
\ss{n-1}{1}{x} \equiv  {\rm Li}_n(x).
\eeq
The $S_{n,\, p}$'s with argument $x$, $1-x$ and $1/x$ can be
related to each other via~\cite{kolbig}
\begin{eqnarray}
\ss{n}{p}{1-x} &=& \sum_{s=0}^{n-1} \frac{\log^s(1-x)}{s!}
\left[
\ss{n-s}{p}{1}
-\sum_{r=0}^{p-1} \frac{(-1)^r}{r!} \log^r(x) ~\ss{p-r}{n-s}{x}\right]
\nonumber \\
&+&  \frac{(-1)^p}{n!\, p!} \log^n(1-x)\log^p(x),\\
\ss{n}{p}{\frac{1}{x}} &=& (-1)^n \sum_{s=0}^{p-1} (-1)^s \sum_{r=0}^{s}
\frac{(-1)^r}{r!} \log^r(-x) \left( \begin{array}{c}n+s-r-1 \\ s-r\end{array}
\right)
~\ss{n+s-r}{p-s}{x} \nonumber \\
&+&
\sum_{r=0}^{n-1} \frac{(-1)^{r+p}}{r!} \log^r(-x)~ C_{n-r,\;p}
+
\frac{(-1)^{n}}{(n+p)!} \log^{n+p}(-x), 
\end{eqnarray}
with
\begin{eqnarray}
C_{n,\, p} &=& (-1)^{n+1} \sum_{r=1}^{p-1} (-1)^{p-r} 
\left( \begin{array}{c}n+r-1 \\ r\end{array}
\right)~
\ss{n+r}{p-r}{-1}
\nonumber \\
&+&  (-1)^p \left(1-(-1)^n \right) ~\ss{n}{p}{-1}.
\end{eqnarray}
The usual binomial coefficient is defined as
\begin{equation}
\left( \begin{array}{c}a \\ b\end{array}
\right) = \frac{a!}{b! \, (a-b)! }.
\end{equation}
We also need the definition of the Riemann Zeta functions 
\beq
\zeta_n= \sum_{s=1}^{\infty}
~\frac{1}{s^n},
\eeq
and in particular
\beq
 \zeta_2 = \frac{\pi^2}{6},  \quad \quad \quad \zeta_3 = 1.20206\ldots
\quad \quad \quad \zeta_4=\frac{\pi^4}{90}.
\eeq

\subsection{Analytic continuation formulae}
\label{subsec:anal}
For $x > 1$, the following analytic continuations should be used
\beqn
\li{x + i0} &=& -\li{\frac{1}{x}} -\frac{1}{2}
 \log^2 (x) +\frac{\pi^2}{3} + i \pi \log (x) \\
\lit{x + i0} &=& \lit{\frac{1}{x}} -\frac{1}{6}\log^3(x) +
\frac{\pi^2}{3}  \log(x) +  \frac{i\pi}{2} \log^2(x) \\ 
\lif{x + i0}  &=& - \lif{\frac{1}{x}}
         -  \frac{1}{24}\log^4(x)  + \frac{\pi^2}{6} \log^2(x)
        + \frac{\pi^4}{45} +
        \frac{i\pi}{6}\log^3(x) \\ 
\ss{1}{2}{x+i0} &=&  -\ss{1}{2}{\frac{1}{x}} + \lit{\frac{1}{x}}  + \log(x)
                       \li{\frac{1}{x}}  + \frac{1}{6} \log^3(x)
                    - \frac{\pi^2}{2}\log(x)  + \zeta_3 \nonumber \\
 && {} + i \pi \left[\frac{\pi^2}{6} - \li{\frac{1}{x}} -
\frac{1}{2}\log^2(x)  \right]    \\
\ss{1}{3}{x+i0} &=& - \ss{1}{3}{\frac{1}{x}} +  \ss{2}{2}{\frac{1}{x}} + 
          \log(x) \ss{1}{2}{\frac{1}{x}} - \lif{\frac{1}{x}} 
       - \log(x) \lit{\frac{1}{x}}
 \nonumber \\
&&  {}
- \frac{1}{2} \log^2(x) \li{\frac{1}{x}} + \frac{\pi^2}{2} \li{\frac{1}{x}}
- \frac{1}{24}\log^4(x) + \frac{\pi^2}{4} \log^2(x) -\frac{19\pi^4}{360}
 \nonumber \\
&&  {}
+i\pi\left[\lit{\frac{1}{x}} - \ss{1}{2}{\frac{1}{x}} 
+\log(x) \li{\frac{1}{x}} + \frac{1}{6} \log^3(x)
- \frac{\pi^2}{6} \log(x) \right] \nonumber \\ \\
\ss{2}{2}{x+i0} &=& \ss{2}{2}{\frac{1}{x}}  -
                      2\lif{\frac{1}{x}} - \log(x) \lit{\frac{1}{x}} 
 + \frac{1}{24}\log^4(x) - \frac{\pi^2}{4}\log^2(x) \nonumber \\
&& {} + \zeta_3 \log(x) 
+\frac{\pi^4}{45} + i \pi \left[ \lit{\frac{1}{x}} -
                      \frac{1}{6}\log^3(x)
+\frac{\pi^2}{6} \log(x) -  \zeta_3 \right].
\eeqn

\subsection{Useful identities}
\label{subsec:ident}

For the expansions of hypergeometric functions given in this paper, 
the argument of the polylogarithms is always either 
$(s+t)/s$ or $(s+t)/t$, which are
related by the following identities
\begin{eqnarray}
\li{\frac{x}{x-1}} &=& -\li{x} - \frac{1}{2}\log^2\(1-x\),\\
\lit{\frac{x}{x-1}} &=& -\lit{x} +\sot{x} + \log\(1-x\)\li{x}
+\frac{1}{6}\log^3\(1-x\),\\
\lif{\frac{x}{x-1}} &=& -\lif{x}+\st2{x}-\so3{x}+\log\(1-x\)\lit{x}
-\log\(1-x\)\sot{x} \nonumber \\
&&-\frac{1}{2}\log^2\(1-x\)\li{x}-\frac{1}{24}\log^4\(1-x\),\\
\sot{\frac{x}{x-1}} &=& \sot{x}-\frac{1}{6}\log^3\(1-x\),\\
\so3{\frac{x}{x-1}} &=& -\so3{x}-\frac{1}{24}\log^4\(1-x\),\\
\st2{\frac{x}{x-1}} &=&
\st2{x}-2\so3{x}-\log\(1-x\)\sot{x}+\frac{1}{24}\log^4\(1-x\).
\end{eqnarray}

\relax
\def\pl#1#2#3{{\it Phys.\ Lett.\ }{\bf #1}\ (#2)\ #3}
\def\zp#1#2#3{{\it Z.\ Phys.\ }{\bf #1}\ (19#2)\ #3}
\def\prl#1#2#3{{\it Phys.\ Rev.\ Lett.\ }{\bf #1}\ (19#2)\ #3}
\def\rmp#1#2#3{{\it Rev.\ Mod.\ Phys.\ }{\bf#1}\ (19#2)\ #3}
\def\prep#1#2#3{{\it Phys.\ Rep.\ }{\bf #1}\ (19#2)\ #3}
\def\pr#1#2#3{{\it Phys.\ Rev.\ }{\bf #1}\ (19#2)\ #3}
\def\np#1#2#3{{\it Nucl.\ Phys.\ }{\bf #1}\ (#2)\ #3}
\def\sjnp#1#2#3{{\it Sov.\ J.\ Nucl.\ Phys.\ }{\bf #1}\ (19#2)\ #3}
\def\app#1#2#3{{\it Acta Phys.\ Polon.\ }{\bf #1}\ (19#2)\ #3}
\def\jmp#1#2#3{{\it J.\ Math.\ Phys.\ }{\bf #1}\ (19#2)\ #3}
\def\jp#1#2#3{{\it J.\ Phys.\ }{\bf #1}\ (19#2)\ #3}
\def\nc#1#2#3{{\it Nuovo Cim.\ }{\bf #1}\ (19#2)\ #3}
\def\lnc#1#2#3{{\it Lett.\ Nuovo Cim.\ }{\bf #1}\ (19#2)\ #3}
\def\ptp#1#2#3{{\it Progr. Theor. Phys.\ }{\bf #1}\ (19#2)\ #3}
\def\tmf#1#2#3{{\it Teor.\ Mat.\ Fiz.\ }{\bf #1}\ (19#2)\ #3}
\def\tmp#1#2#3{{\it Theor.\ Math.\ Phys.\ }{\bf #1}\ (19#2)\ #3}
\def\jhep#1#2#3{{\it J.\ High\ Energy\ Phys.\ }{\bf #1}\ (19#2)\ #3}
\def\epj#1#2#3{{\it Eur.\ Phys. J.\ }{\bf #1}\ (19#2)\ #3}
\relax

\end{document}